\documentclass[a4paper,12pt]{article}
\usepackage{fullpage}
\usepackage{amsfonts,amssymb,amstext,amsmath,amsthm,verbatim,times,cancel,epsfig} 
\usepackage[english]{babel}
\usepackage[T1]{fontenc} 
\usepackage{fancyhdr}
\usepackage{setspace}
\usepackage[english]{varioref}
\usepackage{cite}
\usepackage{bbold}
\usepackage{upgreek}
\usepackage{hyperref}
\hypersetup{colorlinks,linkcolor={black},citecolor={blue},urlcolor={blue}}

\title{Relativistic time-commutative dynamics with $\kappa$-plane noncommutativity}
\author{Alessandro Moia, Stefano Stocchetti, Giovanni Amelino-Camelia\\{\footnotesize\emph{Dipartimento di Fisica ``E. Pancini'', Universit\`a di Napoli ``Federico II'', Via Cinthia 21, 80126 Napoli, Italy}}}
\date{}

\begin{document}
\maketitle

\begin{abstract}
In the last decades, spacetime noncommutativity and the associated deformations of relativistic symmetries have attracted a lot of interest, as several phenomenological windows into quantum gravity are approaching genuine Planck-scale sensitivity. However, the physical significance of the mathematical structures introduced to deal with spacetime noncommutativity is still debated, and some crucial pieces are missing or poorly understood. From time to time it has been suggested that valuable insight into these conceptual challenges could come from the analysis of appropriately designed first-quantized toy models, in which major technical and interpretive issues can be effectively managed or sidestepped altogether. In this paper we take seriously this suggestion and develop the first fully consistent and fully relativistic first-quantized model of two particles propagating and interacting on a noncommutative spacetime. The specific spacetime noncommutativity implemented in our model, which we call `time-commutative $\kappa$-plane', had already been proposed as a suitable arena for a first-quantized analysis, but previous studies were mostly heuristic and failed to provide a full description of the deformed relativistic symmetries. We here go much beyond those pioneering attempts: we find a full characterization of the appropriate deformed Poincaré symmetry algebra as well as its Galilean limit; we build a single-particle quantum model carrying an irreducible representation of the deformed Galilei algebra; we exhibit two consistent, Galilean-relativistic descriptions of a system of two quantum particles interacting via a deformed harmonic potential, finding that the structure of the two-particle symmetry generators is intimately connected with the deformation of the interaction law.
\end{abstract}

\section*{Introduction}\label{introduction}

Over the past decades, the possibility that the relativistic symmetries of spacetime might be deformed, rather than explicitly broken, at some scale $\ell$ of the order of the Planck length has attracted considerable attention within the quantum-gravity community. At ultrashort distances the localization of spacetime events is widely expected to be subject to fundamental limitations arising from the interplay between quantum mechanics and general relativity. However, this does not necessarily imply a failure of the relativity principle, as the current relativistic symmetry group, the Poincaré group, can be modified in such a way as to accommodate a fundamental invariant length $\ell$, in the same way as the former relativistic symmetry group, the Galilei group, was modified so as to accommodate the invariant speed $c$~\cite{qstph}.

Deformed relativistic symmetries have mostly been investigated in the framework of noncommutative geometry, where spacetime coordinates are no longer ordinary commuting numbers but elements of a noncommutative algebra. Spacetime noncommutativity provides a natural way of introducing an intrinsic spacetime delocalization of order $\ell$ without messing with spacetime symmetries. The idea is borrowed from Heisenberg quantum mechanics, where nontrivial commutation rules between positions and conjugate momenta induce an intrinsic phase-space delocalization of order $\hbar$ without impairing the translational nor the rotational invariance of the theory. In such a noncommutative setting, the tools of classical differential geometry become inadequate and a purely algebraic formulation of geometric structures is required. Developments in the theory of quantum groups and Hopf algebras \cite{Hopf algebras for physics at the Planck scale, Drinfeld, foundofquangro, Quantum Groups and Noncommutative Geometry} have provided the mathematical language necessary to extend fundamental geometric notions such as symmetries, homogeneous spaces, and transformations between reference frames to the noncommutative regime.

This new noncommutative framework, however, is still lacking a transparent physical interpretation, making it difficult to derive meaningful phenomenological predictions. In standard quantum theory relativistic symmetries manifest themselves through the first Noether theorem, which guarantees the existence of a set of physical observables, the symmetry generators, whose algebraic relations reflect the Lie structure of the infinitesimal symmetry transformations. The well-known phenomenological signatures of spacetime symmetries, such as the exact conservation of total momentum and angular momentum or the appearance of spectral features associated to irreducible representations of the conserved generators, are all consequences of this crucial result. When the continuous Poincaré group gets deformed into a quantum group, characterized by noncommutative transformation parameters, infinitesimal transformations cease to be well defined, obfuscating the physical import of the deformed symmetries. Even if it is relatively straightforward to generalize the notion of Lie algebra to the noncommutative setting and describe the quantum group in terms of a Hopf algebra of deformed symmetry generators, these mathematical objects cannot be directly associated to any observable quantities, as the noncommutativity of the transformation parameters severs the usual Noether connection between the generators and the algebra of the observables. As a result, it is unclear if and how the formal properties of the deformed generators may translate into actual physical effects.

The most obvious way to overcome these interpretive difficulties would be to find a generalization of Noether theorem to noncommutative spacetime symmetries, through which the deformed generators might be connected to some conserved observable quantities. Several attempts have been made to derive such a connection but, to our knowledge, a fully satisfactory noncommutative-spacetime analogue of Noether theorem is still lacking. Most previous investigations have been framed in the context of Lagrangian field theory on a noncommutative background~\cite{waves, Quantum Field Theory on Noncommutative Spaces}, where standard fields are replaced with functions of the noncommutative coordinates and the deformed generators play the role of generalized differential operators. While formal generalizations of Noether theorem have indeed been obtained for free theories on some noncommutative spacetimes~\cite{Generalizing the Noether theorem for Hopf-algebra spacetime symmetries, nopureboost, fieldtheoryfreidel, Twisted Hopf symmetries of canonical noncommutative spacetimes and the no-pure-boost principle}, the phenomenological relevance of these results is unclear. In standard quantum field theory, fields and the associated observables are not coordinate functions of the (commutative) spacetime manifold, but rather spacetime-indexed operators acting on the appropriate multi-particle Hilbert space. In the noncommutative Lagrangian field theories of Refs. \cite{Generalizing the Noether theorem for Hopf-algebra spacetime symmetries, nopureboost, fieldtheoryfreidel, Twisted Hopf symmetries of canonical noncommutative spacetimes and the no-pure-boost principle}, instead, fields and observables, including the generalized Noether charges, are coordinate functions of the noncommutative spacetime manifold, \emph{i.e.}, elements of the noncommutative algebra generated by the spacetime coordinates. Even if it is still possible to represent the spacetime coordinates, and therefore also the observables, as operators acting on some Hilbert space, it is evident that this representation has nothing to do with the Fock space of standard quantum field theory, undermining the usual physical interpretation in terms of multi-particle states and making it difficult to figure out the phenomenological content of such models. Moreover, Noether-like results obtained so far have been derived in free field theories, where there are no particle reactions and all single-particle momenta, angular momenta, etc. are expected to be separately conserved. In this dynamically trivial context, every function of the single-particle invariants is conserved, and it is therefore impossible to identify with any certainty those few symmetry-protected quantities that would still be conserved in the physical, interacting case. In order to derive genuine conservation laws relevant for particle reactions and phenomenology, one should start from the analysis of interacting models, but it is not at all clear how to manage interactions in noncommutative field theories such as those developed in Refs. \cite{Generalizing the Noether theorem for Hopf-algebra spacetime symmetries, nopureboost, fieldtheoryfreidel, Twisted Hopf symmetries of canonical noncommutative spacetimes and the no-pure-boost principle} and, in any case, there would still be the problem of finding a viable physical interpretation for the resulting observables.

A natural way out of all these difficulties is provided by Galilean-relativistic, first-quantized models of spacetime noncommutativity. Quantum mechanics was originally developed in the Hamiltonian framework of first quantization, in which the classical canonical coordinates $p_i^{I}$ and $x_j^{I}$, $I\in\{1,\dots,N\}$, of an $N$-particle system are promoted to self-adjoint operators $\hat{p}_i^{I}$ and $\hat{x}_j^{I}$ satisfying the Heisenberg commutation rules\footnote{For notational convenience, we work in units such that $\hbar=1$.}
\begin{align}
    [\hat{x}_j^{I},\hat{x}_l^{J}]&=0,   \label{ccr1}\\
    [\hat{p}_i^{I},\hat{x}_j^{J}]&=-i\delta_{ij}\delta^{IJ},   \label{ccr2}\\
    [\hat{p}_i^{I},\hat{p}_k^{J}]&=0,   \label{ccr3}
\end{align}
and observables are identified with self-adjoint elements of the operator algebra generated by $\hat{p}_i^{I}$ and $\hat{x}_j^{I}$. In this Galilean-relativistic context, free-particle dynamics is governed by the Galilei-covariant Hamiltonian $\hat{H}_G=H_G(\hat{p}_i^I)=\sum_I\hat{p}_i^I\hat{p}_i^I/2m_I$ and interactions can be simply described by adding a Galilei-invariant potential term $V(\hat{x}_j^I)$ to $\hat{H}_G$. First quantization seems to be the perfect answer to the interpretive and technical problems of the Lagrangian field-theoretic approach. On the one hand, the particle coordinates $\hat{x}_j^{I}$ are not background indices, but commuting physical operators generating the algebra of the observables. As a result, introducing purely spatial noncommutativity just amounts to appropriately deforming the canonical commutators \eqref{ccr1}-\eqref{ccr3}, and does not require any modifications to the usual physical interpretation of the first-quantized formalism. This interpretation is also quite natural in the light of the original motivation for spacetime noncommutativity, as nontrivial commutation relations among particle coordinates directly translate into intrinsic delocalization (see Ref. \cite{moia} for a more detailed discussion of this point). On the other hand, there is no particular problem with introducing interactions into a first-quantized noncommutative spacetime model: it is sufficient to add an invariant potential term depending on the noncommutative particle coordinates to the (deformed) free Hamiltonian. Any Noether-like result on the deformed Galilei symmetries of such models is thus bound to yield genuine conservation laws, providing crucial insight into the physical manifestations of a deformed relativity principle.

After being suggested in Ref. \cite{soccerball}, the development of a first-quantized, fully dynamical model of spacetime noncommutativity has been first attempted in the recent paper \cite{frattfab}, which focuses on what we shall henceforth call `$\kappa$-plane', \emph{i.e.}, a 2D noncommutative space characterized by the following Lie-algebra commutation rule:
\begin{align}
    [\hat{x}_1,\hat{x}_2]&=i\ell\hat{x}_1.   \label{KappaPlane}
\end{align}
The $\kappa$-plane is the purely spatial version of the much studied 1+1D $\kappa$-Minkowski spacetime \cite{bicross,Classical and Quantum Mechanics of Free  Relativistic Systems}, defined by
\begin{align}
    [\hat{x}_1,\hat{x}_0]&=i\ell\hat{x}_1,   \label{KappaMinkowski}
\end{align}
whose deformed relativistic symmetries are described by the well known 1+1D $\kappa$-Poincaré Hopf algebra \cite{bicross, Classical and Quantum Mechanics of Free  Relativistic Systems, New quantum Poincaré algebra and κ-deformed field theory}. Therefore, it is very easy to obtain a fully consistent Hopf-algebra characterization of the deformed spatial symmetries of the $\kappa$-plane. Starting from this description, the authors of Ref.~\cite{frattfab} build first-quantized models of two and three particles interacting via suitably deformed harmonic potentials, finding out an intriguing connection between the form of the spatial symmetry generators and the deformation of the harmonic interaction.

While clearly showcasing the potential of a first-quantized, fully dynamical approach to spacetime noncommutativity, the study reported in Ref. \cite{frattfab} fails to address the main interpretive issue identified above. In fact, rather than finding a Noether-like connection between the abstract Hopf-algebra generators and some observable operators, the authors simply identify the deformed translation generators $P_i$ with the canonical momenta $\hat{p}_i$, replacing the Heisenberg commutators~\eqref{ccr2} with deformed ones heuristically derived from the Hopf-algebra action of $P_i$ on coordinate products of the form $\hat{x}_jf(\hat{x}_j)$. However, such identification is unacceptable without further justification, since the action $P_i\triangleright f(\hat{x})$ of the translation generators, which does not satisfy the Leibniz rule, cannot be faithfully represented by any deformed commutator $[\hat{p}_i,f(\hat{x})]$. After all, the failure of the Leibniz rule for the action of the deformed generators, which is closely related to the noncommutativity of the transformation parameters, is the very reason why the usual Noether theorem does not apply to noncommutative spacetime symmetries. Other than being undermined by this serious interpretive problem, the exploratory analysis reported in Ref. \cite{frattfab}, being restricted to purely spatial symmetries, is also incomplete from a relativistic standpoint. In order to exhibit a fully Galilean-relativistic, dynamical model of $\kappa$-plane noncommutativity, one should first find the full symmetry algebra, complete with deformed boosts and time translations, of what we shall call `time-commutative $\kappa$-plane', \emph{i.e.}, the 2+1D noncommutative spacetime defined by
\begin{align}
    [\hat{x}_0,\hat{x}_j]&=0,   \label{TCKappaPlane1}\\
    [\hat{x}_1,\hat{x}_2]&=i\ell\hat{x}_1,   \label{TCKappaPlane2}
\end{align}
and then make sure that the first-quantized dynamics is covariant under all deformed Galilei transformations. Simply replacing the term $\hat{p}_i^I\hat{p}_i^I$ in the standard free Hamiltonian $\hat{H}_G$ with the deformed Casimir element of the purely spatial $\kappa$-plane Hopf symmetry algebra, as done in Ref. \cite{frattfab}, can at best be regarded as a heuristic prescription, which, by itself, does not guarantee the proper implementation of a deformed relativity principle.

One of the main difficulties in dealing with deformed boosts and time translations on the time-commutative $\kappa$-plane is that the standard first-quantized formalism is completely non-covariant as regards the time coordinate. While particle spatial coordinates are promoted to self-adjoint observable operators with definite transformation properties under spatial symmetries, time enters the picture as just the transformation parameter associated with time translations. In particular, there is no observable operator that behaves like a particle time coordinate under relativistic transformations. As a result, the action of symmetries involving the time coordinate on the algebra of the observables cannot be directly read off their action on the spacetime manifold, but must rather be indirectly derived from the observable consequences of their group structure. In the case of ordinary Galilei symmetries, this is made possible by the Noether theorem, which guarantees the existence of self-adjoint observables representing (a central extension of) the Galilei Lie algebra. For a single-particle system, which corresponds to an irreducible representation, it is then immediate to identify the canonical momenta with the translation generators and find unique expressions for the other symmetry generators (and the associated unitary transformations) based on their commutation structure (see, \emph{e.g.}, Chapter 3 of Ref. \cite{ballentine}. Unfortunately, in the absence of a generalized Noether theorem, this reconstruction strategy cannot be applied to deformed relativistic symmetries, and it is not clear what one could do about boosts and time translations, at least in the usual, non-covariant first-quantized framework.

In the last decades, it has become clear that single-particle, first-quantized quantum mechanics admits an alternative, manifestly covariant formulation, in which particle spacetime coordinates are treated on the same footing \cite{Trajectories for the wave function of the universe from a simple detector model, Relational time in generally covariant quantum systems: Four models, Spacetime states and covariant quantum theory}. In this covariant quantum mechanics, the time variable $x_0$ and its conjugate momentum $p_0$ are included in an extended set of single-particle canonical coordinates $x_\nu$ and $p_\mu$, which are promoted to self-adjoint operators $\hat{x}_\nu$ and $\hat{p}_\mu$ on a kinematical (\emph{i.e.}, unphysical) Hilbert space and required to satisfy the following commutation relations\footnote{Here and in the following $g_{\mu\nu}$ denotes the Minkowski metric tensor, for which we adopt the ``mostly minuses'' sign convention.}:
\begin{align}
    [\hat{x}_\nu,\hat{x}_\lambda]&=0,   \label{covccr1}\\
    [\hat{p}_\mu,\hat{x}_\nu]&=ig_{\mu\nu},    \label{covccr2}\\
    [\hat{p}_\mu,\hat{p}_\tau]&=0.  \label{covccr3}
\end{align}
Starting from the kinematical Hilbert space and the extended canonical algebra generated by $\hat{p}_\mu$ and $\hat{x}_\nu$, all the physics of the model is then derived by an appropriately chosen self-adjoint Hamiltonian constraint $\hat{C}=C(\hat{p}_\alpha, \hat{x}_\alpha)$ along the lines of Dirac's constrained quantization scheme \cite{quantization of Gauge Systems}. More specifically, the physical Hilbert space of the system is identified requiring that physical states are annihilated by $\hat{C}$, and the observables are obtained as the self-adjoint elements of the extended canonical algebra that commute with $\hat{C}$ (and are therefore well defined on the physical Hilbert space \cite{Refined algebraic quantization: Systems with a single constraint}. Symmetries of the system, including its dynamical evolution, are naturally described in terms of unitary operators commuting with $\hat{C}$, which induce unitary transformations of the physical Hilbert space.

In this covariant framework, the action of the relativistic symmetries on the extended canonical algebra determines their action on the observables, which are built out of the extended canonical variables. Since the canonical variables include a full set $\hat{x}_\nu$ of particle coordinate operators with definite transformation properties under all relativistic symmetries, the difference between boosts, time translations, and purely spatial transformations disappears, and the Hilbert space representation of the whole symmetry group can be directly derived from its action on the spacetime manifold. This is precisely what happens when modeling a free relativistic particle in covariant quantum mechanics \cite{Spacetime states and covariant quantum theory}. In this case, the action of a generic Poincaré transformation $(\Lambda,a)$ on the extended canonical algebra can be written immediately, with no reference to Noether theorem, as
\begin{align}
    \hat{x}_\nu&\mapsto\Lambda_\nu^{\ \alpha}\hat{x}_\alpha+a_\nu,   \label{PoincareX}\\
    \hat{p}_\nu&\mapsto\Lambda_\nu^{\ \alpha}\hat{p}_\alpha.    \label{PoincareP}
\end{align}
Taking the Poincaré-invariant operator
\begin{align}
    \hat{C}_r=\hat{p}_\alpha\hat{p}_\alpha-m^2c^2	\label{Cr}
\end{align}
as the Hamiltonian constraint, the transformations \eqref{PoincareX}-\eqref{PoincareP} are then guaranteed to correspond to genuine quantum symmetries of the algebra of the observables. In particular, their self-adjoint generators $\hat{p}_\mu$ and $\hat{m}_{\rho\sigma}=\hat{x}_\rho\hat{p}_\sigma-\hat{x}_\sigma\hat{p}_\rho$ correspond to well defined observables on the physical Hilbert space, which can be checked to generate the usual spin-0 unitary representation of the Poincaré group. Since this covariant approach does not rely on the Noether theorem for the representation of spacetime transformations on the extended canonical algebra, it is ideally suited to the investigation of the deformed relativistic symmetries of noncommutative spacetimes. It is sufficient to introduce spacetime noncommutativity into the canonical commutation relations \eqref{covccr1}-\eqref{covccr3} and find deformed Poincaré automorphisms of the extended canonical algebra implementing the appropriate quantum group structure. In this way, it is possible to build a fully relativistic, first-quantized model of a single particle propagating on any noncommutative spacetime, and hopefully find a connection between the deformed symmetry group and some observable quantities.

Indeed, in a couple of pioneering papers \cite{fuzzy1, fuzzy2}, this strategy of analysis has been applied to the popular 1+1D $\kappa$-Minkowski spacetime, obtaining very promising results. In Ref. \cite{fuzzy1} the authors show that, by describing the $\kappa$-Poincaré symmetries in terms of a quantum differential calculus \cite{woronowicz}, it is possible to represent the deformed Poincaré generators as linear operators acting on the extended canonical algebra, and the associated noncommutative transformation parameters as functions of the canonical momenta. In this representation, the action $\mathbf{d}\hat{v}$ of a differential $\kappa$-Poincaré transformation on any canonical variable $\hat{v}$ is given by
\begin{align}
    \mathbf{d}\hat{v}=i\varepsilon_\alpha[\hat{p}_\alpha,\hat{v}]+i\beta[\hat{n},\hat{v}],   \label{dvkM}
\end{align}
where $\hat{n}$, like the canonical momenta $\hat{p}_\mu$, is a self-adjoint operator on the kinematical Hilbert space and, most importantly, $\varepsilon_\mu$ and $\beta$ are ordinary, commutative transformation parameters. Expression \eqref{dvkM} is formally identical to the infinitesimal variation associated to a standard Poincaré transformation and suggests that the operators $\hat{p}_\mu$ and $\hat{n}$, which are related, but obviously do not coincide, with the abstract Hopf-algebra translation and boost generators $P_\mu$ and $N$, should be regarded as the true $\kappa$-Poincaré analogues of the usual self-adjoint Poincaré generators. This analogy is substantiated by the fact that, in order for the Hamiltonian constraint $\hat{C}$ to be $\kappa$-Poincaré-invariant, $d\hat{C}=0$, it is necessary and sufficient that $\hat{C}$ commutes with the canonical generators $\hat{p}_\mu$ and $\hat{n}$, meaning that $\hat{p}_\alpha$ and $\hat{n}$ must correspond to well defined observables on the physical Hilbert space, like in the standard case. While restricted to a specific 1+1D $\kappa$-Minkowski single-particle model, this physically transparent Noether-like result provides evidence that full deformed relativistic covariance is associated to the existence of first-quantized observable operators with specific, symmetry-related algebraic properties. In the remainder of Ref. \cite{fuzzy1} and the subsequent paper~\cite{fuzzy2}, the authors investigate the spacetime phenomenology of their model, introducing a novel notion of `fuzzy points', identifying an appropriately deformed position observable of the Newton-Wigner type, and characterizing the spacetime-induced contribution to the associated worldline fuzziness.

The adoption of covariant quantum mechanics in Refs. \cite{fuzzy1, fuzzy2} was mainly motivated by the need of deriving a consistent free-particle phenomenology in the presence of a noncommutative time coordinate, which has no meaning in the standard first-quantized formalism. While convincingly argued, the proposed characterization of the algebra of the observables as a deformed relativistic Heisenberg algebra cannot be regarded as uncontroversial, since the interpretation of the deformed Newton-Wigner operator as a position observable at a given time is at odds with the space-time complementarity seemingly encoded in the $\kappa$-Minkowski commutation rules~\eqref{KappaMinkowski}. Moreover, the focus on empty-space and free-particle $\kappa$-Minkowski phenomenology overshadows what we feel is the most important finding of the study reported in Refs. \cite{fuzzy1, fuzzy2}, namely the discovery of a Noether-like connection between $\kappa$-Poincaré symmetries and first-quantized single-particle observables. In the light of the above discussion, it is clear that generalizing this result to generic spacetime noncommutativity and multi-particle systems with nontrivial dynamics would provide crucial insight into the phenomenology of deformed relativistic symmetries in realistic scenarios.

In the present paper we report substantial progress towards this ambitious goal, exhibiting the first fully consistent and fully Galilean-relativistic first-quantized model of two particles interacting on a noncommutative spacetime. After some preliminaries on Poincaré Hopf algebras and the associated differential calculi (\hyperref[sec: Poincaré Hopf algebras]{Section \ref{sec: Poincaré Hopf algebras}}), we formulate a general notion of first-quantized canonical generators for deformed relativistic symmetries, considerably extending the preliminary results obtained in Ref. \cite{fuzzy1} for 1+1D $\kappa$-Minkowski (\hyperref[sec: Symmetries in CQM]{Section \ref{sec: Symmetries in CQM}}). Then, following Refs.\cite{soccerball,frattfab}, we specialize to the time-commutative $\kappa$-plane~\eqref{TCKappaPlane1}-\eqref{TCKappaPlane2}, finding a complete deformed symmetry algebra at leading order in the noncommutativity parameter $\ell$ (\hyperref[sec: TCK Hopf algebra]{Section \ref{sec: TCK Hopf algebra}}). After introducing $\kappa$-plane noncommutativity into the canonical commutation relations~\eqref{covccr1}-\eqref{covccr3}, we identify the canonical generators associated to the leading-order symmetry algebra found in \hyperref[sec: TCK Hopf algebra]{Section \ref{sec: TCK Hopf algebra}} as well as an appropriate deformation of the Poincaré-invariant constraint~\eqref{Cr} (\hyperref[sec: TCK canonical algebra]{Section \ref{sec: TCK canonical algebra}}). Using standard contraction techniques, we take the Galilean-relativistic limit of our time-commutative $\kappa$-plane model (\hyperref[sec: TCK Galilean limit]{Section \ref{sec: TCK Galilean limit}}) and provide a characterization of the corresponding algebra of the observables as a deformed Heisenberg algebra (\hyperref[sec: TCK particle]{Section \ref{sec: TCK particle}}). Thanks to the commutativity of the time coordinate, we are able to avoid the interpretive difficulties that affect the analogous construction in Ref.\cite{fuzzy1} and thus obtain an ordinary, non-covariant, first-quantized description of a free particle propagating on a $\kappa$-plane background. In particular, we find a deformation of the free Hamiltonian $\hat{p}_i\hat{p}_i/2m$ that is covariant under the full deformed Galilei group, correcting the heuristic result of Ref. \cite{frattfab} and providing it with a rigorous justification as well as a clear physical interpretation. Finally, as a first application of our novel Galilean-relativistic $\kappa$-plane model, we explore the relation between single-particle and two-particle canonical generators in a system of two particles interacting via a deformed harmonic potential (\hyperref[sec: Interacting TCK particles]{Section \ref{sec: Interacting TCK particles}}), corroborating the conclusions reached in Ref. \cite{frattfab} at the heuristic level. Contrary to what happens in the standard case, in which, regardless of the dynamics, the total momentum and angular momentum operators are always given by the sum of the corresponding single-particle generators \cite{jordancurriesudarshan}, we find that our model is compatible with at least two distinct composition laws for the canonical generators, resulting in a peculiar interdependence between the two-particle momentum and the interaction Hamiltonian.

\section{Poincaré Hopf algebras and quantum differential calculi} \label{sec: Poincaré Hopf algebras}

In the next section we generalize the pioneering result of Ref. \cite{fuzzy1} on the representation of deformed relativistic symmetries in covariant quantum mechanics, providing a rigorous definition of canonical, first-quantized generator of a quantum group and stating a Noether-like result of rather broad applicability. As anticipated in the \hyperref[introduction]{Introduction}, our starting point will be the description of deformed relativistic symmetries in terms of a Poincaré Hopf algebra of deformed generators and an associated quantum differential calculus. Here, we briefly recall the basic notions we will need in the following, referring the reader to Refs. \cite{foundofquangro, qgprimer, woronowicz} for a more comprehensive and detailed treatment.

A Poincaré Hopf algebra $\mathcal{P}$ for some (possibly noncommutative) spacetime manifold $\mathcal{X}$ is a unital associative algebra generated by 10 linear operators $P_\mu$ and $M_{\rho\sigma}=-M_{\sigma\rho}$ acting on the coordinate functions of $\mathcal{X}$. The action of the generators $P_\mu$ and $M_{\rho\sigma}$ on a suitably chosen spacetime basis $x_\nu$ is given by
\begin{align}
   P_\mu\triangleright x_\nu&=ig_{\mu\nu},    \label{PAction}\\
   M_{\rho\sigma}\triangleright x_\nu&=i(x_\rho g_{\sigma\nu}-x_\sigma g_{\rho\nu}),  \label{MAction}
\end{align}
and they annihilate the constant function $1(x)=1$:
\begin{equation}
   P_\mu\triangleright1=M_{\rho\sigma}\triangleright1=0.    \label{1Action}
\end{equation}
The action of any Poincaré generator $G\in\{P_\mu,M_{\rho\sigma}\}$ on the product $fh=\mu(f\otimes h)$ of two spacetime functions $f(x)$ and $h(x)$ is completely determined by an algebra homomorphism $\Delta:\mathcal{P}\longrightarrow\mathcal{P}\otimes\mathcal{P}$, called coproduct map, such that
\begin{align}
    G\triangleright fh=G\triangleright\mu(f\otimes h)=\mu(\Delta G\triangleright f\otimes h)=\sum (G_l^{(1)}\triangleright f)(G_l^{(2)}\triangleright h),    \label{GAction}
\end{align}
where we have switched to the more explicit and convenient notation
\begin{equation}
    \Delta G=\sum G_l^{(1)}\otimes G_l^{(2)}.
\end{equation}
In order for \eqref{GAction} to be compatible with the associativity of the coordinate product $\mu$, it is necessary and sufficient that the coproduct be co-associative, \emph{i.e.}, that
\begin{equation}
    \sum\Delta G_l^{(1)}\otimes G_l^{(2)}=\sum G_l^{(1)}\otimes\Delta G_l^{(2)}.    \label{Coassociativity}
\end{equation}
There must also exist an algebra anti-homomorphism $\Sigma:\mathcal{P}\longrightarrow\mathcal{P}$, called antipode map, such that
\begin{equation}
    \sum\Sigma(G_l^{(1)})G_l^{(2)}=\sum G_l^{(1)}\Sigma(G_l^{(2)})=0.  \label{Antipode}
\end{equation}

The fundamental property \eqref{GAction} of the coproduct makes it possible to compute the action of any Poincaré generator $G$, and thus of the whole Hopf algebra $\mathcal{P}$, on every spacetime function starting from the fundamental action laws \eqref{PAction}-\eqref{1Action}. The standard Poincaré Lie algebra $\mathcal{P}_0$ (or rather, its universal enveloping algebra) is the particular Poincaré Hopf algebra obtained by choosing the so called primitive coproduct\footnote{Here and in the following $\mathbb{1}$ will denote the identity operator: $\mathbb{1}\triangleright f=f$.}
\begin{equation}
    \Delta_0G=G\otimes\mathbb{1}+\mathbb{1}\otimes G \label{Delta0}
\end{equation}
and the associated antipode
\begin{equation}
    \Sigma_0(G)=-G, \label{Sigma0}
\end{equation}
which obviously satisfy \eqref{Coassociativity} and \eqref{Antipode}. In this case, the coproduct property \eqref{GAction} reduces to the ordinary Leibniz rule
\begin{equation}
    G\triangleright fh=(G\triangleright f)h+f(G\triangleright h),    \label{LeibnizRule}
\end{equation}
and it is easy to show that the generators admit the usual expressions
\begin{align}
   P_\mu&=ig_{\mu\alpha}\frac{\partial}{\partial x_\alpha},    \label{PDiff}\\
   M_{\rho\sigma}&=i(x_\rho g_{\sigma\alpha}-x_\sigma g_{\rho\alpha})\frac{\partial}{\partial x_\alpha},  \label{MDiff}
\end{align}
as differential operators acting on the commutative coordinate functions of standard Minkowski spacetime. In particular, one can prove that $P_\mu$ and $M_{\rho\sigma}$ satisfy the Poincaré commutation relations
\begin{align}
   [P_\mu,P_\tau]&=0, \label{PPComm}\\
   [P_\mu,M_{\rho\sigma}]&=i(g_{\rho\mu}P_\sigma-g_{\mu\sigma}P_\rho),    \label{PMComm}\\
   [M_{\mu\nu},M_{\rho\sigma}]&=i(g_{\rho\nu}M_{\mu\sigma}-g_{\mu\rho}M_{\nu\sigma}+g_{\sigma\nu}M_{\rho\mu}-g_{\mu\sigma}M_{\rho\nu}).   \label{MMComm}
\end{align}

The primitive coproduct \eqref{Delta0} is generally incompatible with nontrivial spacetime commutation rules of the type
\begin{equation}
    [x_{\nu},x_{\lambda}]=i\ell\Theta_{\nu\lambda}(x),  \label{XX}
\end{equation}
as the action of the Poincaré generators on the right-hand side is clearly dependent on the specific form of $\Theta_{\nu\lambda}(x)$. When dealing with noncommutative spacetimes, it is then necessary to generalize the Leibniz rule by introducing in the fundamental coproducts $\Delta P_\mu$ and $\Delta M_{\rho\sigma}$ (as well as the associated antipodes) non-primitive terms of order $\ell$ that compensate the action of $P_\mu$ and $M_{\rho\sigma}$ on $\Theta_{\nu\lambda}(x)$. Such a deformation makes it impossible to characterize the Poincaré generators as differential operators on a smooth manifold and, most importantly, spoils their Lie-algebra structure, adding nonlinear terms of order $\ell$ to their standard commutators \eqref{PPComm}-\eqref{MMComm}. However, at least at the formal level, the resulting Hopf-algebra structure is rather close to the original Poincaré Lie algebra and provides a natural generalization of the usual geometric description of relativistic symmetries to noncommutative spacetimes.

From a physical standpoint, spacetime symmetries should leave invariant the geometry of the spacetime manifold $\mathcal{X}$ as encoded in the associated algebra of coordinate functions\footnote{In the following, for notational convenience, the symbol $\mathcal{X}$ will also denote this coordinate algebra.}. Technically, this means that any relativistic transformation $T:f\mapsto f'$ should be an algebra homomorphism or, equivalently, that the variation $\mathbf{d}f=f'-f$ should be linear and satisfy the Leibniz rule
\begin{equation}
    \mathbf{d}(fh)=(\mathbf{d}f)h+f(\mathbf{d}h)    \label{dLeibnizRule}
\end{equation}
in the infinitesimal limit $T\to\mathbb{1}$. In the commutative case, it is well known that the infinitesimal variation can be written in terms of $P_\mu$ and $M_{\rho\sigma}$ as
\begin{equation}
    \mathbf{d}f=i\varepsilon_\alpha (P_\alpha\triangleright f)+\frac{i}{2}\omega_{\beta\gamma}(M_{\beta\gamma}\triangleright f)    \label{dfStandard}
\end{equation}
for some real parameters $\varepsilon_\mu$ and $\omega_{\rho\sigma}$. As a result, the action of the Poincaré generators completely encodes the infinitesimal structure of the relativistic symmetry group, providing a full description of the relativistic transformations. In the noncommutative case, equation \eqref{dfStandard} cannot hold, since $P_\mu$ and $M_{\rho\sigma}$ have non-primitive coproducts and do no more satisfy the Leibniz rule. The deformed Poincaré Hopf algebra $\mathcal{P}$ is thus insufficient to completely characterize the relativistic symmetries, and it is necessary to generalize the usual notion of variation.

One way of doing this, already suggested by our notation, is to regard \eqref{dfStandard} as the definition of a first-order differential calculus on standard Minkowski spacetime and deform it so as to accommodate generators with nontrivial coproducts. A generalized (or quantum) first-order differential calculus on a (possibly noncommutative) spacetime manifold $\mathcal{X}$ is a pair $(\mathcal{B},\mathbf{d})$ where $\mathcal{B}$ is a bimodule over $\mathcal{X}$, \emph{i.e.}, an additive group with a compatible left and right multiplication by coordinate functions $f\in\mathcal{X}$, and the quantum differential $\mathbf{d}:\mathcal{X}\longrightarrow\mathcal{B}$ is a linear map satisfying the Leibniz rule \eqref{dLeibnizRule}. The standard variation \eqref{dfStandard} takes values in $\mathcal{X}$ itself, which is obviously an $\mathcal{X}$-bimodule, and can thus be interpreted as defining the action of a (trivial) quantum differential. A straightforward way of generalizing this structure to noncommutative spacetimes is to replace the standard commutative parameters $\varepsilon_\mu$ and $\omega_{\rho\sigma}$ in \eqref{dfStandard} with elements $\upepsilon_\mu$ and $\upomega_{\rho\sigma}$ of some more general $\mathcal{X}$-bimodule $\mathcal{B}$ and write
\begin{equation}
    \mathbf{d}f=i\upepsilon_\alpha(P_\alpha\triangleright f)+\frac{i}{2}\upomega_{\beta\gamma}(M_{\beta\gamma}\triangleright f).    \label{df}
\end{equation}
Requiring that \eqref{df} satisfies the Leibniz rule yields
\begin{multline}
    \upepsilon_\alpha\{(P_\alpha\triangleright fh)-(P_\alpha\triangleright f)h-f(P_\alpha\triangleright h)\}\\
    +\frac{1}{2}\upomega_{\beta\gamma}\{(M_{\beta\gamma}\triangleright fh)-(M_{\beta\gamma}\triangleright f)h-f(M_{\beta\gamma}\triangleright h)\}\\
    =[f,\upepsilon_\alpha](P_\alpha\triangleright h)+\frac{1}{2}[f,\upomega_{\beta\gamma}](M_{\beta\gamma}\triangleright h).    \label{CommPar}
\end{multline}
If $P_\mu$ and $M_{\rho\sigma}$ have primitive coproducts, the generalized transformation parameters $\upepsilon_\mu$ and $\upomega_{\rho\sigma}$ must commute with all spacetime functions, and one recovers the usual notion of infinitesimal variation with $\mathcal{B}=\mathcal{X}$, $\upepsilon_\mu=\varepsilon_\mu1$, and $\upomega_{\rho\sigma}=\omega_{\rho\sigma}1$ for some real $\varepsilon_\mu,\omega_{\rho\sigma}$. Conversely, if $\Delta P_\mu$ and $\Delta M_{\rho\sigma}$ contain non-primitive terms, $\upepsilon_\mu$ and $\upomega_{\rho\sigma}$ must be nontrivial objects that do not commute with spacetime coordinates, and $\mathcal{B}\neq\mathcal{X}$. Noncommutative spacetime symmetries are thus completely characterized by the deformed Poincaré Hopf algebra $\mathcal{P}$ and a compatible quantum differential calculus over the noncommutative manifold $\mathcal{X}$, \emph{i.e.}, a nontrivial $\mathcal{X}$-bimodule $\mathcal{B}$ such that condition \eqref{CommPar} is satisfied for some choice of $\upepsilon_\mu,\upomega_{\rho\sigma}\in\mathcal{B}$. This additional structure on the transformation parameters, which is hidden and does not play any role in the commutative case, has proven crucial in all past attempts at generalizing the Noether theorem to noncommutative spacetime symmetries \cite{Generalizing the Noether theorem for Hopf-algebra spacetime symmetries, nopureboost, fieldtheoryfreidel, Twisted Hopf symmetries of canonical noncommutative spacetimes and the no-pure-boost principle}.

\section{Deformed relativistic symmetries in CQM} \label{sec: Symmetries in CQM}

As first argued in Ref. \cite{fuzzy1} in the particular case of 1+1D $\kappa$-Minkowski noncommutativity, the abstract characterization of deformed relativistic symmetries outlined in the previous section admits a natural and physically transparent representation on the extended canonical algebra $\mathcal{V}$ generated by the canonical operators $\hat{p}_\mu$ and $\hat{x}_\nu$ of covariant quantum mechanics.

In the CQM framework every relativistic symmetry of the system is described by a unitary automorphism $\hat{v}\mapsto\hat{U}\hat{v}\hat{U}^\dagger$ of the extended canonical algebra $\mathcal{V}$ that leaves invariant the Hamiltonian constraint $\hat{C}$. The corresponding unitary operator $\hat{U}$, which must commute with $\hat{C}$, is thus well defined on the physical Hilbert space and induces a unitary automorphism $\hat{o}\mapsto\hat{U}\hat{o}\hat{U}^\dagger$ of the algebra of the observables $\mathcal{O}$, guaranteeing the relativistic covariance of the transition amplitudes. Infinitesimal transformations are implemented by unitary operators of the form $1+i\hat{d}$ infinitesimally close to the identity. In the infinitesimal limit, $\hat{d}$ is a self-adjoint operator commuting with the Hamiltonian constraint, \emph{i.e.}, a physical observable, and the variation $\mathbf{d}\hat{v}$ of any canonical variable $\hat{v}\in\mathcal{V}$ is given by
\begin{equation}
   \mathbf{d}\hat{v}=i[\hat{d},\hat{v}].    \label{dv}
\end{equation}
One thus recovers the connection between infinitesimal symmetries and physical observables familiar from standard quantum mechanics, with the important difference that in CQM equation \eqref{dv} holds on the whole extended canonical algebra $\mathcal{V}$, and not just on the algebra of the observables.

Let $\hat{\mathcal{X}}\subset\mathcal{V}$ denote the sub-algebra generated by the canonical spacetime coordinates $\hat{x}_\nu$, which is a faithful representation of the algebra of the coordinate functions on the (possibly noncommutative) spacetime manifold $\mathcal{X}$. Identifying for simplicity $\hat{\mathcal{X}}$ with $\mathcal{X}$ and setting $\hat{f}=f(\hat{x})\in\mathcal{X}$, the CQM infinitesimal variation
\begin{equation}
   \mathbf{d}\hat{f}=i[\hat{d},\hat{f}]     \label{dhatf}
\end{equation}
defines a quantum differential calculus on $\mathcal{X}$, because it maps $\mathcal{X}$ to $\mathcal{V}$, which is obviously an $\mathcal{X}$-bimodule, and satisfies the Leibniz rule. Since the variation \eqref{dhatf} is induced by a relativistic transformation, the associated differential calculus has to be compatible with the (possibly deformed) spacetime symmetries of $\mathcal{X}$ in the sense of the previous paragraph, \emph{i.e.}, one must have
\begin{equation}
   [\hat{d},\hat{f}]=\hat{\upepsilon}_\alpha(P_\alpha\triangleright\hat{f})+\frac{1}{2}\hat{\upomega}_{\beta\gamma}(M_{\beta\gamma}\triangleright\hat{f})     \label{dcommf}
\end{equation}
for some $\hat{\upepsilon}_\mu,\hat{\upomega}_{\rho\sigma}\in\mathcal{V}$ satisfying condition \eqref{CommPar}. In general, equation \eqref{dcommf} cannot hold if both $\hat{d}$ and $\hat{x}_\nu$ are self-adjoint operators. Since $\hat{d}$ must be a physical observable, this means that the spacetime coordinates $\hat{x}_\nu$ will usually fail to be self-adjoint. Indeed, this is what happens to the $\kappa$-Minkowski time coordinate $\hat{x}_0$ \cite{fuzzy1} as well as the operator $\hat{x}_2$ in our time-commutative model (see \hyperref[sec: TCK canonical algebra]{Section \ref{sec: TCK canonical algebra}}). From a physical point of view, there is no reason to expect, let alone require, that the canonical coordinates of CQM be self-adjoint, as they are just auxiliary mathematical objects used to reconstruct the actual observables. The lack of self-adjointness of $\hat{x}_\nu$ seems strange only because they happen to be self-adjoint in the commutative case. For generic spacetime noncommutativity, the appropriate behaviour of $\hat{x}_\nu$ under adjunction should be determined by making sure that $\hat{d}$ in \eqref{dcommf} is always self-adjoint.

Let us now suppose that one can find dimensionless functions $e_{\alpha\mu}(\ell\hat{p})=\hat{e}_{\alpha\mu}$, $\hat{w}_{\beta\gamma\mu}'$, and $\hat{w}_{\beta\gamma\rho\sigma}$ of the canonical momenta such that
\begin{align}
   \hat{\upepsilon}_\mu&=\varepsilon_\mu\hat{1}+\varepsilon_\alpha\hat{e}_{\alpha\mu}+\frac{\ell}{2}\omega_{\beta\gamma}\hat{w}_{\beta\gamma\mu}',   \label{Epsilon}\\
   \hat{\upomega}_{\rho\sigma}&=\omega_{\rho\sigma}\hat{1}+\omega_{\beta\gamma}\hat{w}_{\beta\gamma\rho\sigma}, \label{Omega}
\end{align}
where $\varepsilon_\mu$ and $\omega_{\rho\sigma}$ are ordinary real parameters. This is clearly possible in the commutative case, for which $\hat{e}_{\alpha\mu}=\hat{w}_{\beta\gamma\mu}'=\hat{w}_{\beta\gamma\rho\sigma}=0$, but it is reasonable to assume that condition \eqref{CommPar}, which is linear in $\upepsilon_\mu$ and $\upomega_{\rho\sigma}$, will also be satisfied by a linear ansatz of the form \eqref{Epsilon}-\eqref{Omega} for several, if not all, noncommutative spacetimes\footnote{Such a representation of the noncommutative transformation parameters is certainly available for the popular $\kappa$-Minkowski spacetime as well as the time-commutative $\kappa$-plane, which is the focus of the following sections. Its full domain of validity will be investigated elsewhere.}. In order for the right-hand side of equation \eqref{dcommf} to depend linearly on $\varepsilon_\mu$ and $\omega_{\rho\sigma}$, this must also be true of the left-hand side or, which is the same, of the operator $\hat{d}$. Hence, there must exist physical observables $\hat{t}_\mu$ and $\hat{m}_{\rho\sigma}=-\hat{m}_{\sigma\rho}$, which we call canonical generators of the (possibly deformed) relativistic symmetry group, such that the CQM infinitesimal variation can be written as
\begin{equation}
   \mathbf{d}\hat{v}=i\varepsilon_\alpha[\hat{t}_\alpha,\hat{v}]+\frac{i}{2}\omega_{\beta\gamma}[\hat{m}_{\beta\gamma},\hat{v}].    \label{dvexplicit}
\end{equation}
Equation \eqref{dvexplicit} generalizes the usual Noether connection between relativistic symmetries and observable quantities to any spacetime allowing a representation of the type \eqref{Epsilon}-\eqref{Omega} for the transformation parameters. The canonical generators are related to the abstract Poincaré generators $P_\mu$ and $M_{\rho\sigma}$ through equation \eqref{dcommf}, which can be rewritten in the more explicit form
\begin{align}
    [\hat{t}_\mu,\hat{f}]&=(P_\mu\triangleright\hat{f})+\hat{e}_{\mu\alpha}(P_\alpha\triangleright\hat{f}),    \label{tcommf}\\
   [\hat{m}_{\rho\sigma},\hat{f}]&=(M_{\rho\sigma}\triangleright\hat{f})+\hat{w}_{\rho\sigma\beta\gamma}(M_{\beta\gamma}\triangleright\hat{f})+\ell\hat{w}_{\rho\sigma\alpha}'(P_\alpha\triangleright\hat{f}). \label{mcommf}
\end{align}

In the commutative case, the operators $\hat{e}_{\alpha\mu}$, $\hat{w}_{\beta\gamma\mu}'$, and $\hat{w}_{\beta\gamma\rho\sigma}$ vanish and \eqref{tcommf}-\eqref{mcommf} reduce to
\begin{align}
    [\hat{t}_\mu,\hat{f}]&=P_\mu\triangleright\hat{f},  \label{tcommf0}\\
   [\hat{m}_{\rho\sigma},\hat{f}]&=M_{\rho\sigma}\triangleright\hat{f}. \label{mcommf0}
\end{align}
Therefore, the canonical generators $\hat{t}_\mu$ and $\hat{m}_{\rho\sigma}$ must generate a self-adjoint representation of the Poincaré Lie algebra $\mathcal{P}_0$ and one recovers the usual statement of the Noether theorem. In particular, $\hat{t}_\mu$ and $\hat{m}_{\rho\sigma}$ must satisfy \eqref{PPComm}-\eqref{MMComm}, and are thus given by $\hat{t}_\mu=\hat{p}_\mu$ and $\hat{m}_{\rho\sigma}=\hat{x}_\rho\hat{p}_\sigma-\hat{x}_\sigma\hat{p}_\rho$, if the particle has no spin; the momentum and angular momentum operators $\hat{p}_i$ and $\hat{m}_{ij}$ commute with the time translation generator $\hat{p}_0$, meaning that the corresponding physical quantities are conserved through the dynamical evolution of the system; and the Hamiltonian constraint $\hat{C}$, which has to commute with all the generators, is naturally identified as the Casimir element \eqref{Cr} of the Poincaré Lie algebra, enforcing the usual relativistic dispersion relation for a particle of mass $m$.

For general spacetimes, the situation is more complex, as the canonical generators do no more generate a representation of the Poincaré Hopf algebra $\mathcal{P}$. By a suitable redefinition of the canonical momenta (see \hyperref[sec: TCK canonical algebra]{Section \ref{sec: TCK canonical algebra}}) one can still have $\hat{t}_\mu=\hat{p}_\mu$, but the expression of the canonical Lorentz generators $\hat{m}_{\rho\sigma}$ is deformed and must be constructively derived from equation \eqref{mcommf}. The resulting commutators $[\hat{p}_\mu,\hat{m}_{\rho\sigma}]$ and $[\hat{m}_{\mu\nu},\hat{m}_{\rho\sigma}]$ can then be obtained through direct computation, and do not generally match the corresponding abstract commutation rules $[P_\mu,M_{\rho\sigma}]$ and $[M_{\mu\nu},M_{\rho\sigma}]$. However, once the canonical generators have been found, one recovers all the physical consequences of the usual Noether theorem: the momentum of the particle is conserved, as the operators $\hat{p}_i$ still commute with the time translation generator $\hat{p}_0$; if the rotation generators $\hat{m}_{ij}$ commute with $\hat{p}_0$, an appropriately deformed notion of angular momentum is conserved too; and the Hamiltonian constraint $\hat{C}$, which determines the relativistic dispersion relation of the particle, can be identified requiring that it commutes with all the deformed Lorentz generators $\hat{m}_{\rho\sigma}$. The only real difference with the commutative case is that the physically relevant algebra is the canonical symmetry algebra $\mathcal{CP}$ generated by $\hat{p}_\mu$ and $\hat{m}_{\rho\sigma}$, not the abstract Poincaré Hopf algebra $\mathcal{P}$. This is not surprising, since $\mathcal{CP}$ encodes the full structure of the infinitesimal relativistic symmetry group, whereas $\mathcal{P}$ is insensitive to the properties of the noncommutative transformation parameters. It seems weird just because, due to the standard Poincaré Lie algebra $\mathcal{P}_0$ being isomorphic to the corresponding $\mathcal{CP}_0$, we are used to identifying the two and formulating the Noether theorem in abstract terms\footnote{Interestingly, this is not true of their Galilean-relativistic limits $\mathcal{G}_0$ and $\mathcal{CG}_0$, which are not isomorphic. Of course this mismatch has nothing to do with noncommutative transformation parameters and does not violate the representation identities \eqref{tcommf0}-\eqref{mcommf0} characteristic of commutative spacetime (see \hyperref[sec: TCK Galilean limit]{Section \ref{sec: TCK Galilean limit}} for a more detailed discussion of this point), but even in this case the physically relevant symmetry algebra is the canonical algebra $\mathcal{CG}_0$, not the abstract Lie algebra $\mathcal{G}_0$.}. In the light of our analysis, though, the availability such a formulation should be regarded as an accidental feature of the commutative limit, at least in the context of first quantization.

\section{Time-commutative-$\kappa$-plane Poincaré Hopf algebra} \label{sec: TCK Hopf algebra}

Following Ref. \cite{frattfab}, we now turn our attention to a specific 2+1D noncommutative spacetime, the time-commutative $\kappa$-plane \eqref{TCKappaPlane1}-\eqref{TCKappaPlane2}. As discussed in the \hyperref[introduction]{Introduction}, this particular spacetime noncommutativity is ideally suited to the development of a Galilean-relativistic first-quantized quantum model with nontrivial interaction. In fact, its spatial sector, the $\kappa$-plane \eqref{KappaPlane}, is the purely spatial version of the 1+1D $\kappa$-Minkowski spacetime \eqref{KappaMinkowski}, whose symmetries are well known, and the commutativity of the time coordinate ensures that the algebra of single-particle observables can be straightforwardly characterized as a deformed Heisenberg algebra in the Galilean-relativistic limit (see \hyperref[sec: TCK particle]{Section \ref{sec: TCK particle}}). Regrettably, a full description of the deformed relativistic symmetries of the time-commutative $\kappa$-plane, as opposed to its spatial sector, is not available in the literature. In this section we fill this gap, exhibiting the time-commutative-$\kappa$-plane (TC$\kappa$) Poincaré Hopf algebra $\mathcal{P}_\kappa$ and characterizing the associated noncommutative transformation parameters at first order in the noncommutativity scale $\ell$. Contrary to most known Poincaré Hopf algebras, which were discovered as abstract objects in the context of research on quantum groups, our first-order TC$\kappa$ algebra $\mathcal{P}_\kappa$ has been found constructively starting from the TC$\kappa$ spacetime commutation rules \eqref{TCKappaPlane1}-\eqref{TCKappaPlane2}. We will thoroughly discuss this novel bottom-up approach to Poincaré Hopf algebras, which can be applied to generic Lie-algebra spacetime noncommutativity, in a forthcoming paper.

Setting for notational convenience $R=M_{12}$ and $N_i=M_{i0}$, the coproducts of the TC$\kappa$ generators are given by
\begin{align} 
    \Delta P_0&=P_0\otimes\mathbb{1}+\mathbb{1}\otimes P_0,  \label{DeltaP0}\\ 
    \Delta P_i&=P_i\otimes\mathbb{1}+\mathbb{1}\otimes P_i+\ell\delta_{i1}P_2\otimes P_1,    \label{DeltaPi}\\
    \Delta R&=R\otimes\mathbb{1}+\mathbb{1}\otimes R+\ell P_2\otimes R,  \label{DeltaR}\\
    \Delta N_i&=N_i\otimes\mathbb{1}+\mathbb{1}\otimes N_i-\ell\epsilon_{ij}P_1\otimes N_j+\ell\delta_{i1}P_0\otimes R,  \label{DeltaNi}
\end{align}
and the corresponding antipodes read
\begin{align}
    \Sigma(P_0)&=-P_0,    \label{SigmaP0}\\
    \Sigma(P_i)&=-P_i+\ell\delta_{i1}P_2P_1, \label{SigmaPi}\\
    \Sigma(R)&=-R+\ell P_2R. \label{SigmaR}\\
    \Sigma(N_i)&=-N_i-\ell\epsilon_{ij}P_1N_j+\ell\delta_{i1}P_0R,   \label{SigmaNi}
\end{align}
It is easy to verify that, at first order in $\ell$, $\Delta$ and $\Sigma$ satisfy the coassociativity and antipode properties \eqref{Coassociativity} and \eqref{Antipode}, and that this Hopf-algebra structure is indeed compatible with the TC$\kappa$ spacetime commutation rules \eqref{TCKappaPlane1}-\eqref{TCKappaPlane2}. The translation and Lorentz commutators $[P_\mu,P_\tau]$ and $[M_{\mu\nu},M_{\rho\sigma}]$ are still given by \eqref{PPComm} and \eqref{MMComm}, which in 2+1D notation read
\begin{align}
    [P_0,P_i]&=[P_i,P_j]=0, \label{PP}\\
    [R,N_i]&=i\epsilon_{ij}N_j,  \label{RNi}\\
    [N_1,N_2]&=-iR. \label{N1N2}
\end{align}
This is not surprising, as it can be argued on general grounds that this feature is common to all physically relevant Poincaré Hopf algebras admitting a homogeneous space \cite{tminkowski}. The mixed commutators $[P_\mu,M_{\rho\sigma}]$, instead, acquire nonlinear contributions of order $\ell$ and are given by
\begin{align}
    [R,P_0]&=0,  \label{RP0}\\
    [R,P_i]&=i\left(\epsilon_{ij}P_j-\frac{\ell}{2}\delta_{i1}P_1^2+\ell \delta_{i1}P_2^2\right),   \label{RPi}\\  [N_i,P_0]&=i\left(P_i-\ell\delta_{i1}P_1P_2+\frac{\ell}{2}\delta_{i2}P_1^2\right), \label{NiP0}\\
    [N_i,P_j]&=i(\delta_{ij}P_0-\ell\epsilon_{ij}P_0P_1+\ell\delta_{i1}\delta_{j1}P_0P_2).  \label{NiPj}
\end{align}
A tedious but straightforward computation shows that, at first order in $\ell$, the commutation rules \eqref{PP}-\eqref{NiPj} are indeed compatible with the homomorphism property of the TC$\kappa$ coproduct, \emph{i.e.}, that
\begin{equation}
    \Delta[G_1,G_2]=[\Delta G_1,\Delta G_2]
\end{equation}
for every $G_1,G_2\in\{P_\mu,M_{\rho\sigma}\}$. The known $\kappa$-plane Hopf algebra, first obtained in Ref.\cite{frattfab} through a Wick rotation of the 1+1D $\kappa$-Poincaré Hopf algebra, is recovered in $\mathcal{P}_\kappa$, at first order in $\ell$, as the spatial sub-algebra\footnote{Ref.\cite{frattfab} adopts a different sign convention for the $\kappa$-plane commutator \eqref{KappaPlane}, resulting in some slight, purely formal differences between their $\kappa$-plane Hopf algebra and ours. The two conventions are related by the replacements $x_2\mapsto-x_2$, $P_2\mapsto-P_2$, and $R\mapsto-R$.} generated by $P_i$ and $R$.

Plugging the TC$\kappa$ coproduct in equation \eqref{CommPar} yields the following nontrivial commutation relations between the noncommutative transformation parameters $\upepsilon_\mu$, $\uptheta=\upomega_{12}$, and $\upbeta_i=\upomega_{i0}$ and the TC$\kappa$ coordinates $x_\nu$ (all other commutators vanish):
\begin{align}
    [\upepsilon_1,x_2]&=i\ell\upepsilon_1,  \label{Epsilonx2}\\
    [\uptheta,x_0]&=-i\ell\upbeta_1,    \label{Thetax0}\\
    [\uptheta,x_2]&=i\ell\uptheta,  \label{Thetax2}\\
    [\upbeta_i,x_1]&=i\ell\epsilon_{ik}\upbeta_k.    \label{Betax1}
\end{align}
It is readily apparent from these equations that one can have $\upbeta_1=0$ if and only if $\upbeta_2=0$, and that $\uptheta$ cannot vanish if $\upbeta_1$ is nonzero. This kind of interdependence among noncommutative transformation parameters was first noticed in the $\kappa$-Minkowski literature, where the impossibility of having $\uptheta=0$ when $\upbeta_i\neq0$ has been framed as a ``no-pure-boost principle'' \cite{nopureboost}. Such an interpretation is indeed unavoidable if one identifies deformed infinitesimal boosts with transformations of the type
\begin{equation}
    \mathbf{d}f=i\upbeta_k(N_k\triangleright f),    \label{FakePureBoost}
\end{equation}
which do not satisfy the Leibniz rule without the addition of a ``deformed infinitesimal rotation'' term $i\uptheta(R\triangleright f)$. This identification, however, is unwarranted, because it is entirely based on a mere formal analogy with the commutative case. What transformations count as pure deformed boosts or rotations in every physical realization of the TC$\kappa$ (or any other) quantum differential calculus will obviously depend on the nature of the noncommutative transformation parameters, which is left unspecified at the abstract level.

In the context of the CQM realization developed in the previous section, a complete characterization of the deformed relativistic symmetries is provided by the canonical generators $\hat{p}_\mu$ and $\hat{m}_{\rho\sigma}$, which reduce to the usual first-quantized symmetry generators in the commutative limit and are associated to unconstrained real parameters $\varepsilon_\mu$ and $\omega_{\rho\sigma}$. It is then natural to identify deformed infinitesimal boosts with transformations of the type
\begin{equation}
    \mathbf{d}\hat{v}=i\beta_k[\hat{n}_k,\hat{v}],  \label{PureBoost}
\end{equation}
where $\hat{n}_i=\hat{m}_{i0}$ and $\beta_i=\omega_{i0}$, which reduce to ordinary boosts in the commutative limit. Contrary to \eqref{FakePureBoost}, expression \eqref{PureBoost} defines a genuine infinitesimal variation on every noncommutative spacetime admitting a canonical symmetry algebra in the sense of the previous section, including the $\kappa$-Minkowski spacetime and the time-commutative $\kappa$-plane. Therefore, pure deformed boosts, as well as pure translations and rotations, always exist and nothing unusual happens in our CQM framework. The puzzling ``no-pure-boost principle'' of Ref. \cite{nopureboost} just reflects the mismatch between canonical and abstract generators encoded in equations \eqref{tcommf}-\eqref{mcommf}, which, in and of itself, is inconsequential at the physical level.

\section{CQM representation of the TC$\kappa$ relativistic symmetries} \label{sec: TCK canonical algebra}

We now introduce TC$\kappa$ spacetime noncommutativity in the extended canonical algebra $\mathcal{V}$ of CQM and identify the TC$\kappa$ canonical symmetry algebra $\mathcal{CP}_\kappa$ by specializing the general representation strategy discussed in \hyperref[sec: Symmetries in CQM]{Section \ref{sec: Symmetries in CQM}} to the time-commutative $\kappa$-plane and its deformed relativistic symmetries as characterized above.

The most convenient way to introduce spacetime noncommutativity into the canonical commutation relations \eqref{covccr1}-\eqref{covccr3} without violating the Jacobi identities, pioneered in Ref. \cite{fuzzy1} for the 1+1D $\kappa$-Minkowski case and discussed in great detail in Ref. \cite{moia} is to start from a standard extended canonical algebra $\mathcal{V}$, generated by ordinary canonical variables $\hat{q}_\nu$ and $\hat{p}_\mu$ satisfying \eqref{covccr1}-\eqref{covccr3}, and look for dimensionless functions $h_{\mu\nu}(\ell\hat{p})=\hat{h}_{\mu\nu}$ and $\hat{k}_{\nu}$ of the canonical momenta such that the deformed spacetime coordinates
\begin{equation}
    \hat{x}_\nu=\hat{q}_\nu+\frac{1}{2}(\hat{q}_\alpha\hat{h}_{\alpha\nu}+\hat{h}_{\alpha\nu}\hat{q}_\alpha)+\ell\hat{k}_\nu    \label{xnu}
\end{equation}
obey the desired commutation rules. One can then forget about the auxiliary commutative coordinates $\hat{q}_\nu$ and regard $\mathcal{V}$ as being generated by deformed canonical variables $\hat{x}_\nu$ and $\hat{p}_\mu$ satisfying
\begin{align}
    [\hat{x}_\nu,\hat{x}_\lambda]&=i\ell\Theta_{\nu\lambda}(\hat{x}),   \label{defccr1}\\
    [\hat{p}_\mu,\hat{x}_\nu]&=ig_{\mu\alpha}(\delta_{\alpha\nu}\hat{1}+\hat{h}_{\alpha\nu}),    \label{defccr2}\\
    [\hat{p}_\mu,\hat{p}_\tau]&=0.  \label{defccr3}
\end{align}
It can be shown that, given a deformed canonical algebra of the form \eqref{defccr1}-\eqref{defccr3}, with $\Theta_{\nu\lambda}(\hat{x})$ and $\hat{h}_{\mu\nu}$ satisfying the appropriate Jacobi identities, it is always possible to choose $\hat{k}_\nu$ so that the coordinates $\hat{q}_\nu$, implicitly defined by \eqref{xnu}, and $\hat{p}_\mu$ obey the undeformed commutation rules \eqref{covccr1}-\eqref{covccr3} \cite{moia}. This means that every possible CQM model of spacetime noncommutativity\footnote{In principle, one could obtain an even more general CQM implementation of \eqref{defccr1} by relaxing the commutativity condition \eqref{defccr3} on the momenta. However, we are here studying flat-spacetime scenarios where gravity is negligible and deformed translations commute, and we are about to identify $\hat{p}_\mu$ with the canonical translation generators $\hat{t}_\mu$. Therefore, there would be no point in contemplating noncommutative momenta.} can be obtained through a representation of the form \eqref{xnu} for the noncommutative coordinates $\hat{x}_\nu$ and is completely specified by the choice of $\hat{h}_{\mu\nu}$ and $\hat{k}_\nu$. These two operators are constrained, but not determined by the spacetime commutation relations \eqref{defccr1}. For Lie-algebra spacetimes, like the $\kappa$-Minkowski spacetime and the time-commutative $\kappa$-plane, one can always set $\hat{k}_\nu=0$ \cite{moia}, but the choice of $\hat{h}_{\mu\nu}$, which determines the deformed Heisenberg commutators \eqref{defccr2}, is not a matter of convention and should be physically motivated.

The form of $\hat{h}_{\mu\nu}$ is intimately related to the definition of the momentum operators. In fact, given a set of canonical momenta $\hat{p}_\nu$ satisfying \eqref{defccr2}-\eqref{defccr3}, one can obtain another set satisfying \eqref{defccr2}-\eqref{defccr3} via the nonlinear redefinition
\begin{equation}
    \hat{p}_\mu\mapsto(\delta_{\mu\alpha}\hat{1}+\hat{l}_{\mu\alpha})\hat{p}_\alpha, \label{redefp}
\end{equation}
where $\hat{l}_{\mu\tau}=l_{\mu\tau}(\ell\hat{p})$. It is clear that, by appropriately varying $\hat{l}_{\mu\tau}$, one can obtain every possible $\hat{h}_{\mu\nu}$ in \eqref{defccr2}, and that choosing a specific $\hat{h}_{\mu\nu}$ is equivalent to fixing the definition of the canonical momenta. In the commutative case, as discussed in \hyperref[sec: Symmetries in CQM]{Section \ref{sec: Symmetries in CQM}}, one has $\hat{t}_\mu=\hat{p}_\mu$, so that momenta are physically identified by the requirement that they are the canonical generators of spacetime translations. In a generic noncommutative model, the translation generators $\hat{t}_\mu$ may not coincide with the momenta. However, they are always of the form $\hat{t}_\mu=(\delta_{\mu\alpha}\hat{1}+\hat{l}_{\mu\alpha})\hat{p}_\alpha$ for some $\hat{l}_{\mu\tau}$, as they must reduce to $\hat{p}_\mu$ in the commutative limit and satisfy
\begin{equation}
    [\hat{t}_\mu,\hat{x}_\nu]=ig_{\alpha\nu}(\delta_{\mu\alpha}\hat{1}+\hat{e}_{\mu\alpha}).    \label{tcommx}
\end{equation}
It is then natural to exploit the freedom \eqref{redefp} in the definition of the canonical momenta to enforce the usual identification $\hat{t}_\mu=\hat{p}_\mu$ and have the deformed Heisenberg relations \eqref{defccr2} encode the action of the deformed translation generators on the noncommutative coordinates $\hat{x}_\nu$. With this redefinition, $\hat{h}_{\mu\nu}$ is given by $g_{\mu\alpha}\hat{e}_{\alpha\beta}g_{\beta\nu}$ and can be immediately read off the representation \eqref{Epsilon} of the noncommutative translation parameters.

While the canonical momenta $\hat{p}_\mu$, once identified with the translation generators, must always be self-adjoint operators, regardless of the specific deformation of the canonical commutation rules, the properties of $\hat{x}_\nu$ or, which is the same, $\hat{q}_\nu$ under adjunction cannot be prescribed \emph{a priori}. Rather, as observed in \hyperref[sec: Symmetries in CQM]{Section \ref{sec: Symmetries in CQM}}, they are to be derived in any particular model from the requirement that the deformed Lorentz generators $\hat{m}_{\rho\sigma}$ are self-adjoint. However, in order to do this, it is often convenient to parametrize the lack of self-adjointness of $\hat{q}_\nu$ in terms of a further dimensionless function $a(\ell\hat{p})=\hat{a}$ of the momenta via the redefinition
\begin{equation}
    \hat{q}_\nu\mapsto(\hat{1}+\hat{a})^{-1}\hat{q}_\nu(\hat{1}+\hat{a})=\hat{q}_\nu+(\hat{1}+\hat{a})^{-1}[\hat{q}_\nu,\hat{a}], \label{redefq}
\end{equation}
which does not affect the canonical commutation relations \eqref{covccr1}-\eqref{covccr3}. In this way it is always possible to have the commutative coordinates $\hat{q}_\nu$ be self-adjoint in exchange for just a slight complication of the representation \eqref{xnu} of the deformed spacetime coordinates $\hat{x}_\nu$.

We are now ready to specialize to the TC$\kappa$ case and identify the associated canonical symmetry algebra $\mathcal{CP}_\kappa$, working as before at first order in the noncommutativity scale $\ell$. Substituting the general formula \eqref{Epsilon} for $\upepsilon_\mu$ in \eqref{Epsilonx2} yields the unique solution $\hat{e}_{\alpha\mu}=-\ell\delta_{\alpha1}\delta_{\mu1}p_2$ and $\hat{w}_{\beta\gamma\mu}'=0$. The TC$\kappa$ noncommutative coordinates are thus given by
\begin{equation}
    \hat{x}_\nu=\hat{q}_\nu-\ell\delta_{\nu1}\hat{p}_2\hat{q}_1,    \label{TCKxnu}
\end{equation}
and satisfy the deformed Heisenberg commutation rules
\begin{equation}
    [\hat{p}_\mu,\hat{x}_\nu]=i(g_{\mu\nu}\hat{1}-\ell g_{\mu1}\delta_{\nu1}p_2).   \label{TCKpx}
\end{equation}
At first order in $\ell$, redefining the commutative coordinates $\hat{q}_\nu$ via \eqref{redefq} just amounts to adding as many imaginary constants:
\begin{equation}
    \hat{q}_\nu\mapsto\hat{q}_\nu+i\ell a_\nu\hat{1}. 
\end{equation}
It is thus possible to rewrite the TC$\kappa$ coordinates as
\begin{equation}
    \hat{x}_\nu=\hat{q}_\nu+i\ell a_\nu\hat{1}-\ell\delta_{\nu1}\hat{p}_2\hat{q}_1,    \label{TCKxnu+a}
\end{equation}
where $\hat{q}_\nu$ are self-adjoint and the imaginary terms, to be determined \emph{a posteriori}, guarantee the self-adjointness of the deformed Lorentz generators $\hat{m}_{\rho\sigma}$.

Setting $\vartheta=\omega_{12}$ and $\beta_{i}=\omega_{i0}$ in the general formula \eqref{Omega} for $\upomega_{\rho\sigma}$ and plugging it in \eqref{Thetax0}-\eqref{Betax1} yields the following representation of the noncommutative parameters $\uptheta$ and $\upbeta_i$:
\begin{align}
    \uptheta&=\vartheta\hat{1}-\ell\beta_1\hat{p}_0-\ell\vartheta\hat{p}_2, \label{TCKTheta}\\
    \upbeta_i&=\beta_i\hat{1}-\ell\epsilon_{ik}\beta_k\hat{p}_1.    \label{TCKBeta}
\end{align}
The action of the Lorentz generators $\hat{m}_{\rho\sigma}$ on $\hat{x}_\nu$ is then given by \eqref{mcommf}, which in 2+1D notation ($\hat{r}=\hat{m}_{12}$, $\hat{n}_i=\hat{m}_{i0}$) reads
\begin{align}
    [\hat{r},\hat{x}_0]&=0, \label{TCKrx0}\\
    [\hat{r},\hat{x}_j]&=i(\hat{1}-\ell\hat{p}_2)\epsilon_{jk}\hat{x}_k, \label{TCKrxj}\\
    [\hat{n}_i,\hat{x}_0]&=i\hat{x}_i+i\ell\epsilon_{ik}\hat{p}_1\hat{x}_k,  \label{TCKnix0}\\
    [\hat{n}_i,\hat{x}_j]&=i(\delta_{ij}\hat{1}+\ell\epsilon_{ij}\hat{p}_1)\hat{x}_0-i\ell\delta_{i1}\epsilon_{jk}\hat{p}_0\hat{x}_k. \label{TCKnixj}
\end{align}
These relations, once rewritten in terms of the self-adjoint commutative coordinates $\hat{q}_\nu$, make it possible to univocally identify the generators $\hat{r}$ and $\hat{n}_i$ as well as fix the still unspecified constants $a_\nu$. Substituting the expression \eqref{TCKxnu+a} for $\hat{x}_\nu$ in \eqref{TCKrx0}-\eqref{TCKnixj} yields
\begin{align}
    [\hat{r},\hat{q}_0]&=0, \label{TCKrq0}\\
    [\hat{r},\hat{q}_j]&=i\epsilon_{jk}\hat{q}_k-i\ell\delta_{j1}\left(\hat{q}_1\circ\hat{p}_1+\frac{1}{2}[\hat{p}_1,\hat{q}_1]-ia_2\right)+2i\ell\delta_{j2}\left(\hat{q}_1\hat{p}_2-\frac{i}{2}a_1\right), \label{TCKrqj}\\
    [\hat{n}_i,\hat{q}_0]&=i\hat{q}_i-i\ell\delta_{i1}(\hat{q}_1\hat{p}_2-\hat{q}_2\hat{p}_1-ia_1)-i\ell\delta_{i2}\left(\hat{q}_1\circ\hat{p}_1+\frac{1}{2}[\hat{p}_1,\hat{q}_1]-ia_2\right),  \label{TCKniq0}\\
    [\hat{n}_i,\hat{q}_j]&=i\delta_{ij}(\hat{q}_0+i\ell a_0)-i\ell\delta_{j1}\epsilon_{ik}(\hat{q}_k\hat{p}_0-\hat{q}_0\hat{p}_k)+i\ell\delta_{j2}\delta_{i1}(\hat{q}_1\hat{p}_0+\hat{q}_0\hat{p}_1), \label{TCKniqj}
\end{align}
where
\begin{equation}
    \hat{v}_1\circ\hat{v}_2=\frac{1}{2}(\hat{v}_1\hat{v}_2+\hat{v}_2\hat{v}_1)  \label{SymmProd}
\end{equation}
denotes the symmetrized product between two canonical variables $\hat{v}_1,\hat{v}_2\in\mathcal{V}$. Taking the adjoint of equations \eqref{TCKrq0}-\eqref{TCKniqj} it is easy to see that $\hat{r}$ and $\hat{n}_i$ are self-adjoint if and only if $a_0=a_1=0$ and $a_2=-1/2$. With this choice of $a_\nu$, as anticipated in \hyperref[sec: Symmetries in CQM]{Section \ref{sec: Symmetries in CQM}}, $\hat{x}_2=\hat{q}_2-i\ell/2$ fails to be self-adjoint and the unique solution of \eqref{TCKrq0}-\eqref{TCKniqj} reads
\begin{align}
    \hat{r}&=\hat{q}_1\hat{p}_2-\hat{q}_2\hat{p}_1+\frac{\ell}{2}\hat{p}_1\hat{q}_1\hat{p}_1-\ell\hat{p}_2^2\hat{q}_1,  \label{TCKr}\\
    \hat{n}_1&=\hat{q}_1\hat{p}_0-\hat{q}_0\hat{p}_1+\ell\hat{p}_0\hat{p}_1\hat{q}_2-\ell\hat{p}_1\hat{p}_2\hat{q}_0-\ell\hat{p}_2\hat{p}_0\hat{q}_1,   \label{TCKn1}\\
    \hat{n}_2&=\hat{q}_2\hat{p}_0-\hat{q}_0\hat{p}_2+\frac{\ell}{2}\hat{p}_1^2\hat{q}_0-\ell\hat{p}_0\hat{p}_1\circ\hat{q}_1.    \label{TCKn2}
\end{align}

The commutation structure of the TC$\kappa$ canonical algebra $\mathcal{CP}_\kappa$ can now be obtained by explicit computation. The translation and Lorentz sectors, as in the TC$\kappa$ Hopf algebra $\mathcal{P}_\kappa$, are undeformed:
\begin{align}
    [\hat{p}_0,\hat{p_i}]&=[\hat{p}_i,\hat{p}_j]=0, \label{TCKpp}\\
    [\hat{r},\hat{n}_i]&=i\epsilon_{ij}\hat{n}_j,   \label{TCKrni}\\ 
    [\hat{n}_1,\hat{n}_2]&=-i\hat{r},   \label{TCKn1n2}
\end{align}
whereas the mixed commutators are given by
\begin{align}
    [\hat{r},\hat{p}_0]&=0, \label{TCKrp0}\\
    [\hat{r},\hat{p}_i]&=i\left(\epsilon_{ik}\hat{p}_k+\frac{\ell}{2}\delta_{i1}\hat{p}_1^2-\ell\delta_{i1}\hat{p}_2^2\right),   \label{TCKrpi}\\
    [\hat{n}_i,\hat{p}_0]&=i\left(\hat{p}_i+\ell\delta_{i1}\hat{p}_1\hat{p}_2-\frac{\ell}{2}\delta_{i2}\hat{p}_1^2\right),  \label{TCKnip0}\\
    [\hat{n}_i,\hat{p}_j]&=i(\delta_{ij}\hat{p}_0+\ell\epsilon_{ij}\hat{p}_0\hat{p}_1-\ell\delta_{i1}\delta_{j1}\hat{p}_0\hat{p}_2).  \label{TCKnipj}
\end{align}
These are similar, but not identical, to the commutation rules \eqref{RP0}-\eqref{NiPj} characterizing $\mathcal{P}_\kappa$, due to a sign reversal of the nonlinear corrections. Interestingly, the same reversal phenomenon occurs in 1+1D $\kappa$-Minkowski at first order in $\ell$ \cite{fuzzy1}. In any case, as already argued in \hyperref[sec: Symmetries in CQM]{Section \ref{sec: Symmetries in CQM}}, it is the algebraic properties of $\mathcal{CP}_\kappa$, and not those of $\mathcal{P}_\kappa$, that are observationally relevant in the light of our generalized Noether theorem. Since $\mathcal{CP}_\kappa$ is a sub-algebra of the TC$\kappa$ algebra of the observables $\mathcal{O}_\kappa$, covariance under TC$\kappa$ deformed relativistic symmetries ensures the existence of self-adjoint operators $\hat{p}_\mu,\hat{r},\hat{n}_i\in\mathcal{O}_\kappa$ satisfying the commutation relations \eqref{TCKpp}-\eqref{TCKnipj}. In particular, the TC$\kappa$ deformed spatial momenta $\hat{p}_i$ and angular momentum $\hat{r}$, like their standard counterparts, must commute with the time translation generator $\hat{p}_0$, meaning that the corresponding physical quantities must be conserved through the dynamical evolution of the system. Moreover, the operator
\begin{equation}
    \hat{C}_\kappa=\hat{p}_\alpha\hat{p}_\alpha-m^2c^2-\ell\hat{p}_1^2\hat{p}_2,  \label{Ckappa}
\end{equation}
which, at first order in $\ell$, is the only function of the canonical momenta that commutes with the whole $\mathcal{CP}_\kappa$ and reduces to the standard Poincaré invariant \eqref{Cr} in the commutative limit, is the only natural choice for the Hamiltonian constraint, resulting in a distinctive TC$\kappa$ deformation of the relativistic dispersion relation. It is important to emphasize that the construction carried out in this section has no further physical content. While it could be tempting to attach some physical meaning to the expressions \eqref{TCKr}-\eqref{TCKn2} for the Lorentz generators in terms of the canonical variables, they were strictly instrumental to deriving the algebraic structure of $\mathcal{CP}_\kappa$ and do not encode any other physical properties of the model. In particular, they do not provide a representation of the Lorentz generators in terms of spacetime observables, as the spacetime variables $\hat{q}_\nu$ (or $\hat{x}_\nu$), which do not commute with $\hat{C}_\kappa$, are not observable. Deformed spatial observables $\hat{x}_j(0)$ akin to the Heisenberg position operators of standard quantum mechanics will be identified in \hyperref[sec: TCK particle]{Section \ref{sec: TCK particle}}, where the Galilean-relativistic limit of $\mathcal{CP}_\kappa$ will be characterized as a deformed Heisenberg algebra and the Galilei generators will be written as functions of the canonical observables $\hat{p}_i$ and $\hat{x}_j(0)$.

Before closing this section, it is instructive to invert relations \eqref{tcommf}-\eqref{mcommf} so as to obtain an explicit representation of the TC$\kappa$ abstract symmetry generators $P_\mu$, $N_i$, and $R$ as linear operators on the spacetime coordinate algebra $\mathcal{X}\subset\mathcal{V}$. Since $\hat{e}_{\mu\alpha}=-\ell\delta_{\mu1}\delta_{\alpha1}\hat{p}_2$, equation \eqref{tcommf} can be immediately inverted and reads
\begin{equation}
    P_\mu\triangleright\hat{f}=[\hat{p}_\mu,\hat{f}]+\ell\delta_{\mu1}\hat{p}_2[\hat{p}_1,\hat{f}].    \label{Pactf}
\end{equation}
An analogous, albeit less straightforward, inversion of equation \eqref{mcommf} starting from the representation \eqref{TCKTheta}-\eqref{TCKBeta} of the noncommutative transformation parameters $\uptheta$ and $\upbeta_i$ yields
\begin{align}
   R\triangleright\hat{f}&=(\hat{1}+\ell\hat{p}_2)[\hat{r},\hat{f}],    \label{Ractf}\\
   N_i\triangleright\hat{f}&=[\hat{n}_i,\hat{f}]-\ell\epsilon_{ik}\hat{p}_1[\hat{n}_k,\hat{f}]+\ell\delta_{i1}\hat{p}_0[\hat{r},\hat{f}].   \label{Niactf}
\end{align}
Even if the action laws \eqref{Ractf}-\eqref{Niactf} involve left multiplication by functions of the canonical momenta, it can be verified by explicit computation that they do indeed map $\mathcal{X}$ to $\mathcal{X}$, and that the operators $P_\mu$, $R$, and $N_i$ satisfy all the algebraic and co-algebraic properties of the TC$\kappa$ Hopf algebra $\mathcal{P}_\kappa$ characterized in \hyperref[sec: TCK Hopf algebra]{Section \ref{sec: TCK Hopf algebra}}. In particular, the momentum-dependent factors are responsible for the violation of the Leibniz rule and the appearance of the non-primitive terms of the TC$\kappa$ coproducts \eqref{DeltaP0}-\eqref{DeltaNi}. Expressions \eqref{Ractf}-\eqref{Niactf} provide an explicit illustration of the mismatch between abstract and canonical symmetry generators pointed out at the end of the previous section when discussing the ``no-pure-boost principle'' of Ref. \cite{nopureboost}. In order for an infinitesimal transformation of the type
\begin{equation}
    \mathbf{d}\hat{f}=i\uptheta(R\triangleright\hat{f})+i\upbeta_k(N_k\triangleright\hat{f})
\end{equation}
to satisfy the Leibniz rule, \emph{i.e.}, to describe a genuine relativistic symmetry, it is necessary that all the terms in \eqref{Ractf}-\eqref{Niactf} which are not linear in the canonical generators cancel out. And the only way of getting rid of the quadratic term $\ell\delta_{i1}\hat{p}_0[\hat{r},\hat{f}]$ in the action law of $N_1$ is by appropriately adjusting the analogous order-$\ell$ contribution coming from $\uptheta(R\triangleright\hat{f})$, meaning that ``pure boosts'' of the form \eqref{FakePureBoost} cannot be actual symmetries. Far from proving that pure deformed boosts of the time-commutative $\kappa$-plane do not exist, this ``no-pure-boost'' phenomenon just reflects the fact that, at the physical level, the abstract deformed symmetry generators are rather complicated objects with no transparent interpretation, as clearly evidenced by equations \eqref{Pactf}-\eqref{Niactf}.

\section{TC$\kappa$ symmetries in the Galilean-relativistic limit} \label{sec: TCK Galilean limit}

Having derived the TC$\kappa$ canonical symmetry algebra $\mathcal{CP}_\kappa$, we could now study its irreducible representations and build a fully relativistic model of a free quantum particle propagating on a TC$\kappa$ background. By identifying the appropriate spacetime observables, we would then be able to investigate the associated spacetime phenomenology along the lines of Ref. \cite{fuzzy2}, where the worldline fuzziness associated to spacetime noncommutativity was first characterized in the 1+1D $\kappa$-Minkowski case. However, these interesting developments will be pursued elsewhere. As anticipated in the \hyperref[introduction]{Introduction}, we are here interested in building first-quantized models with nontrivial interaction, and this is only possible in the Galilean-relativistic limit, since Poincaré-covariant dynamics requires mediating fields and is notoriously incompatible with a first-quantized, potential-based description (see, \emph{e.g.}, the ample discussion in Chapter 1 of Ref. \cite{strocchiqft}). Therefore, in this section, we study the TC$\kappa$ symmetry algebras $\mathcal{CP}_\kappa$ and $\mathcal{P}_\kappa$ in the low-speed limit $c\to\infty$, obtaining a full characterization of the corresponding deformed Galilei symmetry algebras $\mathcal{CG}_\kappa$ and $\mathcal{G}_\kappa^G$. These results will then be used in \hyperref[sec: TCK particle]{Section \ref{sec: TCK particle}} and \hyperref[sec: Interacting TCK particles]{Section \ref{sec: Interacting TCK particles}} to characterize the single-particle algebra of the observables in the Galilean-relativistic limit and build the first ever Galilean-relativistic model of two particles interacting on a TC$\kappa$ background.

The most convenient and physically transparent way of taking the low-speed limit of the TC$\kappa$ Poincaré canonical algebra $\mathcal{CP}_\kappa$ is to rewrite the infinitesimal variation \eqref{dvexplicit} as
\begin{equation}
   \mathbf{d}\hat{v}=i\tau[\hat{h},\hat{v}]+i\varepsilon_k[\hat{p}_k,\hat{v}]+i\vartheta[\hat{r},\hat{v}]+i\upsilon_l[\hat{k}_l,\hat{v}],    \label{dvG}
\end{equation}
where the new time translation and boost parameters $\tau=\varepsilon_0/c$ and $\upsilon_i=\beta_ic$ have dimensions of time and speed, respectively, and the corresponding rescaled generators are given by\footnote{Technically, the subtraction of the rest energy contribution $mc^2\hat{1}$ from the rescaled time translation generator $\hat{p}_0c$ is unjustified at this stage and should be determined \emph{a posteriori} requiring that the Hamiltonian constraint $\hat{C}$ stay finite in the limit $c\to\infty$. However, this step is not affected by spacetime noncommutativity and the right subtraction is well known from the analysis of the Galilean-relativistic limit in the standard, commutative case.}
\begin{align}
    \hat{h}&=c\hat{p}_0-mc^2\hat{1},    \label{h}\\
    \hat{k}_i&=\frac{1}{c}\hat{n}_i.     \label{k}
\end{align}
In the Galilean-relativistic limit, one has that $c\to\infty$, $\varepsilon_0\to\infty$, and $\beta_i\to0$, but both $\tau$ and $\upsilon_i$ remain constant. As a result, when $c\to\infty$, \eqref{dvG} reduces to the Galilei variation and the rescaled generators $\hat{h}$, $\hat{p}_i$, $\hat{r}$, and $\hat{k}_i$ can be identified with the Galilei generators. The Galilei commutators can thus be straightforwardly obtained by taking the formal limit $c\to\infty$ in the commutators of the rescaled Poincaré generators. Rewriting the undeformed translation and Lorentz commutators \eqref{TCKpp}-\eqref{TCKn1n2} in terms of $\hat{h}$ and $\hat{k}_i$ one finds
\begin{align}
    [\hat{h},\hat{p_i}]&=[\hat{p}_i,\hat{p}_j]=0, \label{TCKhp}\\
    [\hat{r},\hat{k}_i]&=i\epsilon_{ij}\hat{k}_j,   \label{TCKrki}\\ 
    [\hat{k}_1,\hat{k}_2]&=-\frac{i}{c^2}\hat{r},   \label{TCKk1k2}
\end{align}
which for $c\to\infty$ reduce to the undeformed Galilei commutators
\begin{align}
    [\hat{h},\hat{p_i}]&=[\hat{p}_i,\hat{p}_j]=0, \label{GTCKhp}\\
    [\hat{r},\hat{k}_i]&=i\epsilon_{ij}\hat{k}_j,   \label{GTCKrki}\\ 
    [\hat{k}_1,\hat{k}_2]&=0.   \label{GTCKk1k2}
\end{align}
An analogous treatment of the TC$\kappa$ mixed commutators \eqref{TCKrp0}-\eqref{TCKnipj} yields the following deformed Galilei commutators:
\begin{align}
    [\hat{r},\hat{h}]&=0, \label{GTCKrp0}\\
    [\hat{r},\hat{p}_i]&=i\left(\epsilon_{ik}\hat{p}_k+\frac{\ell}{2}\delta_{i1}\hat{p}_1^2-\ell\delta_{i1}\hat{p}_2^2\right),   \label{GTCKrpi}\\
    [\hat{k}_i,\hat{h}]&=i\left(\hat{p}_i+\ell\delta_{i1}\hat{p}_1\hat{p}_2-\frac{\ell}{2}\delta_{i2}\hat{p}_1^2\right),  \label{GTCKnip0}\\
    [\hat{k}_i,\hat{p}_j]&=i(\delta_{ij}m\hat{1}+\ell\epsilon_{ij}m\hat{p}_1-\ell\delta_{i1}\delta_{j1}m\hat{p}_2).  \label{GTCKnipj}
\end{align}
Relations \eqref{GTCKhp}-\eqref{GTCKnipj} define the TC$\kappa$ Galilei canonical algebra $\mathcal{CG}_\kappa$, which is easily checked to correctly reduce to the standard Galilei canonical algebra $\mathcal{CG}_0$ in the commutative limit. The associated Galilei-invariant Hamiltonian constraint, at first order in $\ell$, is given by the Galilean-relativistic limit of \eqref{Ckappa}, which reads
\begin{equation}
    \hat{C}_\kappa=2m\hat{h}-\hat{p}_k\hat{p}_k-\ell\hat{p}_1^2\hat{p}_2.  \label{GCkappa}
\end{equation}
The TC$\kappa$ deformation of the relativistic dispersion relation predictably translates into a deformation of the Galilean-relativistic Hamiltonian operator (see \hyperref[sec: TCK particle]{Section \ref{sec: TCK particle}}).

The abstract Galilei Hopf algebra $\mathcal{G}_\kappa$ can be derived in an analogous manner. The infinitesimal variation \eqref{df} can be rewritten as
\begin{equation}
   \mathbf{d}f=i\uptau(H\triangleright f)+i\upepsilon_k(P_k\triangleright f)+i\uptheta(R\triangleright f)+i\upupsilon_l(K_l\triangleright f),    \label{dfG}
\end{equation}
where $\uptau=\upepsilon_0/c$, $\upupsilon_i=\upbeta_ic$, $H=cP_0$, and $K_i=N_i/c$. In the Galilean-relativistic limit, $\uptau$ and $\upupsilon_i$ remain constant, meaning that \eqref{dfG} reduces to the Galilei variation and $H$, $P_i$, $R$, and $K_i$ can be identified with the abstract Galilei generators. Rewriting the fundamental coproducts \eqref{DeltaP0}-\eqref{DeltaNi} in terms of the rescaled generators and taking the limit $c\to\infty$ yields
\begin{align} 
    \Delta H&=H\otimes\mathbb{1}+\mathbb{1}\otimes H,  \label{DeltaH}\\ 
    \Delta P_i&=P_i\otimes\mathbb{1}+\mathbb{1}\otimes P_i+\ell\delta_{i1}P_2\otimes P_1,    \label{GDeltaPi}\\
    \Delta R&=R\otimes\mathbb{1}+\mathbb{1}\otimes R+\ell P_2\otimes R,  \label{GDeltaR}\\
    \Delta K_i&=K_i\otimes\mathbb{1}+\mathbb{1}\otimes K_i-\ell\epsilon_{ij}P_1\otimes K_j,  \label{DeltaKi}
\end{align}
and the corresponding antipodes are given by
\begin{align}
    \Sigma(H)&=-H,    \label{SigmaH}\\
    \Sigma(P_i)&=-P_i+\ell\delta_{i1}P_2P_1, \label{GSigmaPi}\\
    \Sigma(R)&=-R+\ell P_2R. \label{GSigmaR}\\
    \Sigma(K_i)&=-K_i-\ell\epsilon_{ij}P_1K_j.   \label{SigmaKi}
\end{align}
The translation and Lorentz commutators, as in the canonical algebra $\mathcal{CG}_\kappa$, reduce to the undeformed Galilei commutators
\begin{align}
    [H,P_i]&=[P_i,P_j]=0, \label{HP}\\
    [R,K_i]&=i\epsilon_{ij}K_j,  \label{RKi}\\
    [K_1,K_2]&=0, \label{K1K2}
\end{align}
while the mixed commutators are given by
\begin{align}
    [R,H]&=0,  \label{RH}\\
    [R,P_i]&=i\left(\epsilon_{ij}P_j-\frac{\ell}{2}\delta_{i1}P_1^2+\ell \delta_{i1}P_2^2\right),   \label{GRPi}\\  [K_i,H]&=i\left(P_i-\ell\delta_{i1}P_1P_2+\frac{\ell}{2}\delta_{i2}P_1^2\right), \label{KiH}\\
    [K_i,P_j]&=0.  \label{KiPj}
\end{align}

In addition to the sign reversal of the nonlinear order-$\ell$ corrections, already noticed above, there is another striking structural difference between the two TC$\kappa$ Galilei symmetry algebras $\mathcal{G}_\kappa$ and $\mathcal{CG}_\kappa$: the abstract generators $K_i$ and $P_j$ commute, whereas the commutators $[\hat{k}_i,\hat{p}_j]$ are nontrivial and proportional to the mass $m$ of the particle. This phenomenon is the noncommutative generalization of the well known mismatch between the standard Galilei Lie algebra $\mathcal{G}_0$ and its canonical counterpart $\mathcal{CG}_0$. In fact, the standard canonical algebra $\mathcal{CG}_0$ is a central extension, not a true representation, of the Galilei Lie algebra $\mathcal{G}_0$. This is possible because equations \eqref{tcommf0}-\eqref{mcommf0}, as well as their Galilei analogues, are insensitive to the addition of constant multiples of the identity to the commutation relations among the canonical generators. In the Poincaré case the additional constants can all be removed by a suitable redefinition of the generators, but in the Galilei case no such adjustment can reabsorb a term of the form $ib\delta_{ij}\hat{1}$ in the commutator $[\hat{k}_i,\hat{p}_j]$, leaving open the possibility that $b\neq0$. As a matter of fact, the physically relevant canonical algebra $\mathcal{CG}_0$ describing the Galilei symmetries of ordinary quantum mechanics has $b=m$, because in the real world relativistic symmetries are actually described by the Poincaré canonical algebra $\mathcal{CP}_0$, and $\mathcal{CP}_0$, as shown above, reduces to $\mathcal{CG}_0$, not to $\mathcal{G}_0$, in the low-speed limit. In the presence of spacetime noncommutativity, the underdetermination of the Galilei canonical algebra gets more severe, as further multiples of the identity can be introduced in the representation of the noncommutative Galilei transformation parameters, leading to the appearance of additional non-constant terms in the commutation relations among the canonical generators. In principle, given a noncommutative spacetime, one could study which terms can or cannot be reabsorbed via suitable redefinitions and characterize all possible deformed Galilei canonical algebras, but this is a rather challenging mathematical problem and one would still have to identify the physical algebra. In order to do physics in the Galilean-relativistic limit, it is much easier to start from the relevant Poincaré canonical algebra, which is not affected by any underdetermination, and take the low-speed limit to obtain directly the corresponding Galilei canonical algebra, as we did above in the TC$\kappa$ case. This explains why in the previous sections we focused on the characterization of the TC$\kappa$ Poincaré symmetry algebras, even if we are chiefly interested in the Galilean-relativistic regime. The abstract Galilei Hopf algebra $\mathcal{G}_\kappa$, while not directly relevant to the implementation of our Galilean-relativistic quantum model, will be instrumental to guessing the form of the two-particle symmetry generators in one of the two scenarios analyzed in \hyperref[sec: Interacting TCK particles]{Section \ref{sec: Interacting TCK particles}}.

\section{Free Galilean-relativistic particle on a TC$\kappa$ background}    \label{sec: TCK particle}

Knowing $\mathcal{CG}_\kappa$, we are now ready to study Galilean-relativistic quantum mechanics on a TC$\kappa$ background. Even if both the general characterization of deformed relativistic symmetries and the derivation of $\mathcal{CG}_\kappa$ reported in the previous sections would have been impossible to obtain without being framed in CQM terms, the following discussion of the physics of a free TC$\kappa$ quantum particle, as well as the two-particle analysis of \hyperref[sec: Interacting TCK particles]{Section \ref{sec: Interacting TCK particles}}, will be formulated in the standard, non-covariant language of ordinary quantum mechanics. The reason is that, once the TC$\kappa$ deformed algebra of the observables has been identified, it can be directly characterized on the physical Hilbert space with standard representation techniques, without passing through the kinematical degrees of freedom of CQM. In this way, it is easier to identify the deformed Heisenberg position observables and provide the whole construction with a transparent physical interpretation, firmly grounded in standard quantum mechanical intuition.

In relativistic quantum mechanics, free elementary particles are described in terms of irreducible representations of the relevant canonical symmetry algebra, namely $\mathcal{CP}_0$ or $\mathcal{CG}_0$. The idea, originating with Wigner, is that elementary systems should not contain parts and should thus be described by the smallest possible Hilbert spaces allowing the implementation of the relativistic symmetries. Since this argument is completely insensitive to spacetime noncommutativity, it is natural to model a free Galilean-relativistic particle propagating on a TC$\kappa$ background in terms of an irreducible representation of the TC$\kappa$ canonical algebra $\mathcal{CG}_\kappa$. In an irreducible representation, the TC$\kappa$ Galilei algebra $\mathcal{CG}_\kappa$ is not just contained, but actually coincides with the algebra of the observables $\mathcal{O}_\kappa$. Every single-particle observable $o$ must thus be associated to a self-adjoint function $\hat{o}=o(\hat{h},\hat{p}_i,\hat{r},\hat{k}_j)$ of the deformed Galilei generators. Moreover, not all the symmetry generators are independent, because Schur's lemma requires that all the operators commuting with the whole symmetry algebra $\mathcal{CG}_\kappa$ be constant multiples of the identity. One such operator is the Galilei Hamiltonian constraint \eqref{GCkappa} identified in the previous section, which must vanish if the particle satisfies the TC$\kappa$ deformed dispersion relation. The deformed Galilean-relativistic Hamiltonian operator $\hat{h}$, describing the dynamical evolution of a free TC$\kappa$ particle, can thus be written in terms of the spatial momenta as 
\begin{equation} 
    \hat{h}=\frac{1}{2m}(\hat{p}_k\hat{p}_k+\ell\hat{p}_1^2\hat{p}_2).  \label{hirr}
\end{equation}
This expression correctly reduces to the usual free Hamiltonian in the commutative limit and is indeed proportional, as heuristically suggested in Ref. \cite{frattfab}, to the spatial invariant $\hat{p}_k\hat{p}_k+\ell\hat{p}_1^2\hat{p}_2$, \emph{i.e.}, the Casimir element of the deformed Euclidean algebra generated by $\hat{p}_i$ and $\hat{r}$. However, the relevant Euclidean algebra is the canonical, not the abstract one, whose spatial invariant, due to the sign reversal of the order-$\ell$ corrections, reads $P_kP_k-\ell P_1^2P_2$. The authors of Ref. \cite{frattfab} made the right heuristic guess but confused the canonical and the abstract symmetry algebras, getting the right result with a wrong sign\footnote{The fact that equation (14) in Ref. \cite{frattfab} formally agrees with our \eqref{hirr} is an accident due to our different sign conventions for the $\kappa$-plane commutator \eqref{KappaPlane}. In our variables, obtained from those of Ref. \cite{frattfab} through the replacements $x_2\mapsto-x_2$, $P_2\mapsto-P_2$, and $R\mapsto-R$, the cubic correction to the free Hamiltonian reported in Ref.\cite{frattfab} has a negative sign.}. This incident perfectly illustrates the importance of going beyond heuristics and not relying on commutative physical intuition when dealing with spacetime noncommutativity and deformed relativistic symmetries. Other than the Galilei Hamiltonian constraint \eqref{GCkappa}, it can be verified that the self-adjoint operator
\begin{equation}
    \hat{W}=m\hat{r}-\hat{p}_2\hat{k}_1+\hat{p}_1\hat{k}_2+\ell\hat{p}_1\hat{p}_2\hat{k}_2+\frac{\ell}{2}\hat{p}_1^2\hat{k}_1,  \label{W}
\end{equation}
which is a deformation of the Galilean-relativistic spin operator, commutes with the whole canonical algebra $\mathcal{CG}_\kappa$ at first order in $\ell$. For a spin-0 particle\footnote{Here and in the following we disregard spin, as it is trivial and uninteresting in our 2+1D toy model. This simplification will have to be reconsidered in the analysis of more realistic 3+1D scenarios.} $\hat{W}$ must vanish, meaning that the angular momentum operator $\hat{r}$ is given by the following function of $\hat{p}_i$ and $\hat{k}_j$:
\begin{equation}
    \hat{r}=\frac{1}{m}\hat{p}_2\hat{k}_1-\frac{1}{m}\hat{p}_1\hat{k}_2-\frac{\ell}{m}\hat{p}_1\hat{p}_2\hat{k}_2-\frac{\ell}{2m}\hat{p}_1^2\hat{k}_1.  \label{rirr}
\end{equation}
Therefore, the TC$\kappa$ single-particle algebra of the observables $\mathcal{O}_\kappa$ is actually generated by four independent self-adjoint operators $\hat{p}_i$ and $\hat{k}_j$ satisfying
\begin{align}
    [\hat{p}_i,\hat{p}_j]&=0.   \label{ppirr}\\
    [\hat{p}_i,\hat{k}_j]&=-i(\delta_{ij}m\hat{1}-\ell\epsilon_{ij}m\hat{p}_1-\ell\delta_{i1}\delta_{j1}m\hat{p}_2), \label{pkirr}\\
    [\hat{k}_i,\hat{k}_j]&=0,   \label{kkirr}
\end{align}

Looking at the quasi-canonical commutation relations \eqref{ppirr}-\eqref{kkirr}, it is apparent that $\mathcal{O}_\kappa$ is a deformed Heisenberg algebra. However, its physical interpretation is not straightforward due to the boost generators $\hat{k}_j$ not corresponding to any meaningful observable quantity. In standard commutative quantum mechanics, the usual interpretation of the Heisenberg algebra is obtained identifying the operators $\hat{k}_j/m$ with $\hat{x}_j(0)$, the Heisenberg initial-time position observables, based on the known transformation properties of position vectors under Galilei symmetries \cite{ballentine}. For a TC$\kappa$ particle this identification is unwarranted, as the operators $\hat{k}_i/m$ still commute with each other, while any TC$\kappa$ position observables should obey the $\kappa$-plane commutation rule
\begin{equation}
    [\hat{x}_1(0),\hat{x}_2(0)]=i\ell\hat{x}_1(0).  \label{x0x0}
\end{equation}
Nonetheless, it is reasonable to assume that there exist deformed position observables $\hat{x}_j(0)\in\mathcal{O}_\kappa$ satisfying \eqref{x0x0} and reducing to the standard Heisenberg operators in the commutative limit. These deformed position operators should transform like $\kappa$-plane coordinates under TC$\kappa$ deformed spatial translations and rotations, \emph{i.e.}, satisfy the deformed Heisenberg commutation rules
\begin{equation}
    [\hat{p}_i,\hat{x}_j(0)]=-i\delta_{ij}\hat{1}+i\ell\delta_{i1}\delta_{j1}\hat{p}_2,  \label{px0}
\end{equation}
as well as the symmetrized version of equation \eqref{TCKrxj}:
\begin{equation}
    [\hat{r},\hat{x}_j(0)]=i\epsilon_{jk}\hat{x}_k(0)-i\ell\epsilon_{jk}\hat{p}_2\circ\hat{x}_k(0). \label{rx0}
\end{equation}
At first order in $\ell$, it is easy to verify that
\begin{equation}
    \hat{x}_j(0)=\frac{1}{m}\hat{k}_j-\frac{\ell}{m}\epsilon_{jk}\hat{k}_k\circ\hat{p}_1 \label{xj0}
\end{equation}
are the only self-adjoint $\ell$-deformations of the standard Heisenberg position operators obeying \eqref{px0}, and that they also satisfy both \eqref{x0x0} and \eqref{rx0}. It is then natural to identify $\hat{x}_j(0)$ as TC$\kappa$ initial-time position operators and regard the algebra of the observables $\mathcal{O}_\kappa$ as a deformed Heisenberg algebra generated by $\hat{p}_i$ and $\hat{x}_j(0)$. In terms of the new, physically meaningful variables, the fundamental commutation relations \eqref{ppirr}-\eqref{kkirr} read
\begin{align}
    [\hat{p}_i,\hat{p}_j]&=0.   \label{TCKccr1}\\
    [\hat{p}_i,\hat{x}_j(0)]&=-i\delta_{ij}\hat{1}+i\ell\delta_{i1}\delta_{j1}\hat{p}_2, \label{TCKccr2}\\
    [\hat{x}_i(0),\hat{x}_j(0)]&=i\ell\epsilon_{ij}\hat{x}_1(0),   \label{TCKccr3}
\end{align}
the canonical boost generators are given by
\begin{equation}
    \hat{k}_j=m\hat{x}_j(0)+\ell m\epsilon_{jk}\hat{x}_k(0)\circ\hat{p}_1,  \label{kirr}
\end{equation}
and the angular momentum operator \eqref{rirr} can be rewritten in the more transparent form
\begin{equation}
    \hat{r}=\hat{p}_2\hat{x}_1(0)-\hat{p}_1\hat{x}_2(0)+\frac{\ell}{2}\hat{p}_1\hat{x}_1(0)\hat{p}_1.  \label{rirr2}
\end{equation}

Equations \eqref{TCKccr1}-\eqref{rirr2}, together with expression \eqref{hirr} for the deformed free-particle Hamiltonian, are the TC$\kappa$ analogues of the fundamental formulas of ordinary quantum mechanics and provide an almost complete physical description of a free quantum particle propagating on a TC$\kappa$ background in the Galilean-relativistic regime. The only missing ingredient is a characterization of the quantum states of the particle, \emph{i.e.}, the identification of a physical Hilbert space carrying an irreducible representation of the independent observables $\hat{p}_i$ and $\hat{x}_j(0)$. The easiest way of doing this is to adapt the method employed in \hyperref[sec: TCK canonical algebra]{Section \ref{sec: TCK canonical algebra}} to reconstruct the extended canonical algebra and write the deformed position operators $\hat{x}_j(0)$ as
\begin{equation}
    \hat{x}_j(0)=\hat{q}_j(0)-\ell\delta_{j1}\hat{p}_2\hat{q}_1(0).   \label{qj0}
\end{equation}
It can be immediately verified that the spatial momenta $\hat{p}_i$ and the self-adjoint auxiliary operators $\hat{q}_j(0)$ implicitly defined by \eqref{qj0} satisfy the standard canonical commutation relations \eqref{ccr1}-\eqref{ccr3}, meaning that $\mathcal{O}_\kappa$ can be regarded as an ordinary Heisenberg algebra generated by $\hat{p}_i$ and $\hat{q}_j(0)$. But it is well known \cite{strocchiqm} that every irreducible representation of the Heisenberg algebra, under very mild regularity conditions, is unitarily equivalent to the standard momentum representation, in which physical states are described by normalizable wavefunctions $\psi(p)\in L^2(\mathbb{R}^2,d^2p)$ and the canonical coordinates $\hat{p}_i$ and $\hat{q}_j(0)$ act as multiplication and derivation operators:
\begin{align}
    \hat{p}_i\psi(p)&=p_i\psi(p),   \label{ppsi}\\
    \hat{q}_j(0)\psi(p)&=i\frac{\partial\psi(p)}{\partial p_j}. \label{q0psi}
\end{align}
As a result, the quantum state of a TC$\kappa$ particle can be characterized in terms of an ordinary momentum wavefunction $\psi(p)$, and the deformed position observables act as the differential operators
\begin{equation}
    \hat{x}_j(0)\psi(p)=i\frac{\partial\psi(p)}{\partial p_j}-i\ell\delta_{j1}p_2\frac{\partial\psi(p)}{\partial p_1}.   \label{x0psi}
\end{equation}
With the explicit implementation of the canonical observables as operators on a physical Hilbert space, the description of our TC$\kappa$ Galilean-relativistic free-particle model is complete. Starting from \eqref{ppsi} and \eqref{x0psi}, every observable of interest can be realized as an operator on $L^2(\mathbb{R}^2,d^2p)$, making it possible to compute spectra, expectation values, and transition amplitudes. For example, one could characterize the intrinsic spatial delocalization encoded in the $\kappa$-plane commutation rule \eqref{x0x0} in terms of a Heisenberg-like uncertainty principle for the $x_1$ and $x_2$ coordinates of the particle, or compute the spacetime-induced contribution to wave-packet spreading at first order in the noncommutativity scale $\ell$. However, we postpone the investigation of TC$\kappa$ free-particle phenomenology, as we are here mainly interested in the application of our Galilean-relativistic first-quantized framework to the development of noncommutative spacetime models with nontrivial interaction. In the next section, in order to illustrate the potential of our approach in the characterization of interacting scenarios, we build the first-ever Galilean-relativistic model of two TC$\kappa$ particles interacting through a deformed harmonic potential, corroborating the preliminary heuristic results of Ref.~\cite{frattfab} on the identification of the two-particle generators and paving the way for a more comprehensive analysis of this crucial issue.

\section{Galilean-relativistic interacting TC$\kappa$ particles} \label{sec: Interacting TCK particles}

As a first application of the Galilean-relativistic $\kappa$-plane framework, we study a two-particle system with deformed harmonic interaction, addressing the long-standing issue of the composition laws of canonical generators in DSR-relativistic theories. We show that the model allows multiple consistent composition laws, provided that the interaction potential is suitably deformed, thereby placing the results of Ref. \cite{frattfab} on firmer theoretical ground.

The algebra of observables for the composite system, $\mathcal{O}^{\mathrm{tot}}_\kappa$, is naturally defined as the tensor product of two copies of the single-particle algebra $\mathcal{O}_\kappa$ introduced in the previous section,
\begin{equation}
\mathcal{O}^{\mathrm{tot}}_\kappa=\mathcal{O}^{1}_\kappa \otimes \mathcal{O}^{2}_\kappa.
\end{equation}
To realize TC$\kappa$-Galilei covariance at the level of the composite system, one must identify the two-particle generators in $\mathcal{O}^{\mathrm{tot}}_\kappa$ that satisfy the canonical algebra $\mathcal{CG}_\kappa$ \eqref{GTCKhp}-\eqref{GTCKnipj} \emph{i.e.}, one has to construct a reducible representation of this algebra. The relation between the single-particle and two-particle generators then defines the generators composition law.
 Jordan et al. in Ref. \cite{jordancurriesudarshan}  showed that in Special Relativity the composition law for every generator is simple addition. More specifically, starting from the Poincaré algebra and the equations defining the transformation properties of the particle positions under the Poincaré group, they showed that the symmetry generators of the composite system are just the sum of the corresponding single-particle generators. By contrast, applying the same procedure to the Galilei group reveals that the generator of time translations for the composite system is not uniquely fixed, but may be modified by the addition of a Galilei-invariant term, which is conventionally interpreted as the interaction potential. 
 In a DSR-relativistic theory a rigorous characterization of this kind is still lacking. In fact, when one generalizes symmetries to a Hopf algebraic setup, things get significantly more complicated, since the linear composition law ceases to be a viable option. Let us specialize to our CQM model of the TC$\kappa$-plane. It is evident that the composition law cannot be of the form
\begin{eqnarray}
    &\hat{P}_i= \sum_I \hat{p}_i^I, \\
    & \hat{R}= \sum_I \hat{r}^I, \\
    & \hat{K}_i= \sum_I \hat{k}_i^I, \\
    &\hat{H}= \sum_I \hat{h}^I
\end{eqnarray}
as would not preserve the non-linear commutation relations of the algebra \eqref{GTCKhp}-\eqref{GTCKnipj}. In fact, the linearity of the commutation relation of the Poincaré algebra is a crucial element for the linearity of the composition law.

Since the non-interacting case provides no information on the form of possible deformations of the composition law, the main objective of Ref.~\cite{frattfab} was to study the dependence of the composition law on the form of the interaction among particles; more specifically, they related the composition law with possible deformations of a very common and important potential, the harmonic potential. In this study, which was limited to the spatial sector of the algebra of symmetries, two possible deformations of the composition law were proposed: the first, labelled "proper-dS composition law", is motivated by arguments emerging from the curved momentum space literature~\cite{desitter}
\begin{align} \label{frattfabDS1}
    &\hat{P}_1=\hat{p}_1^A+\hat{p}_1^B-\ell(\hat{p}_2^A\hat{p}_1^B+\hat{p}_1^A\hat{p}_2^B),\\ \label{frattfabDS2}
    &\hat{P}_2=\hat{p}_2^A+\hat{p}_2^B+\ell\hat{p}_1^A\hat{p}_1^B,\\ \label{frattfabDS3}
    &\hat{R}=\hat{r}^A+\hat{r}^B.
\end{align}
Indeed, one can show that with this composition law momentum space acquires the geometrical structure of de Sitter space. The second, inspired by the form of the coproduct of the corresponding generators, reads
\begin{align} \label{frattfabCOP1}
    &\hat{P}_1=\hat{p}_1^A+\hat{p}_1^B-\ell\hat{p}_2^A\hat{p}_1^B,\\ \label{frattfabCOP2}
    &\hat{P}_2=\hat{p}_2^A+\hat{p}_2^B,\\ \label{frattfabCOP3}
    &\hat{R}=\hat{r}^A+\hat{r}^B-\ell\hat{p}_2^A\hat{r}^B. 
\end{align}

We are now going to reproduce the results of Ref.~\cite{frattfab}, but in the fully rigorous CQM framework we have set up, and we shall extend it to the full algebra of relativistic symmetries, working as usual at first order in the noncommutativity scale $\ell$. In particular, we will confirm that the composition law is closely related with the form of the dynamical interaction.
As in Ref. \cite{frattfab}, we shall focus on deformations of the harmonic oscillator hamiltonian 
\begin{align}
   \hat{H}^{AB}_0=\frac{(\hat{p}_1^A)^2}{2m^A}+\frac{(\hat{p}_2^A)^2}{2m^A}+\frac{(\hat{p}_1^B)^2}{2m^B}+\frac{(\hat{p}_2^B)^2}{2m^B}+\frac{1}{2}g(\hat{\vec{x}}^A(0)-\hat{\vec{x}}^B(0))^2,
\end{align}
where $g$ is the coupling constant, the labels $A$ and $B$ refer to the two particles interacting. The kinetic term will be deformed according to \eqref{hirr}
\begin{align} 
    \hat{H}_{KIN}^{AB}=\frac{(\hat{p}_1^A)^2}{2m^A}+\frac{(\hat{p}_2^A)^2}{2m^A}+\ell\frac{(\hat{p}_1^A)^2\hat{p}_2^A}{2m^A}+\frac{(\hat{p}_1^B)^2}{2m^B}+\frac{(\hat{p}_2^B)^2}{2m^B}+\ell\frac{(\hat{p}_1^B)^2\hat{p}_2^B}{2m^B}.
\end{align}
We parametrize the deformation of the harmonic potential as
\begin{equation} \label{eq: 5.94}
    V^{AB}=V(\hat{\vec{x}}^A(0),\hat{\vec{x}}^B(0))=\frac{1}{2}g(\hat{\vec{x}}^A(0)-\hat{\vec{x}}^B(0))^2+\ell g( \alpha_{ijk}^{IJK}\hat{p}_i^I\circ\hat{x}_j^J(0)\circ\hat{x}_k^K(0)+\beta_l^L\hat{x}_l^L(0)),
\end{equation}
where $\alpha_{ijk}^{IJK}$ and $\beta_l^L$ are numerical coefficients, the sum extends both to spatial indices (lowercase $i$, $j$, $k$, and $l$ letters) and particle indices (uppercase letters), and
\begin{equation}
    \hat{o}_1\circ\hat{o}_2\circ\hat{o}_3=\frac{1}{6}(\hat{o}_1\hat{o}_2\hat{o}_3+\hat{o}_1\hat{o}_3\hat{o}_2+\hat{o}_2\hat{o}_1\hat{o}_3+\hat{o}_2\hat{o}_3\hat{o}_1+\hat{o}_3\hat{o}_1\hat{o}_2+\hat{o}_3\hat{o}_2\hat{o}_1)  \label{SymmProd3}
\end{equation}
denotes the symmetrized product between three observables.
First, we investigate the ``proper-dS composition law'' mentioned above.
Our first concern is to extend \eqref{frattfabDS1}-\eqref{frattfabDS3} with the boost generator and the time translation generator. Since our algebra \eqref{GTCKhp}-\eqref{GTCKnipj} has an undeformed Lorentz sector we make the ansatz
\begin{align}
    &\hat{K}_i=\hat{k}_i^A+\hat{k}_i^B, \\
    &\hat{H}=\hat{h}^A+\hat{h}^B.
\end{align}
Indeed, we can check the compatibility of this composition law  with the algebra \eqref{GTCKhp}-\eqref{GTCKnipj}:
\begin{align} \label{algebracomposite1}
    &[\hat{R},\hat{P}_i]=i(\epsilon_{ij}\hat{P}_j+\frac{\ell}{2}\delta_{i1}\delta_{j1}\hat{P}_j^2-\ell \delta_{1i}\delta_{2j}\hat{P}_j^2),\\
    &[\hat{K}_i,\hat{H}]=i(\hat{P}_i+\ell\delta_{1i}\delta_{j2}\delta_{1k}\hat{P}_j\hat{P}_k-\frac{\ell}{2}\delta_{i2}\delta_{j1}\hat{P}_j^2), \\
    &[\hat{R},\hat{K}_i]=i\epsilon_{ij}\hat{K}_j,\\ \label{algebracomposite2}
    &[\hat{K}_i,\hat{P}_j]=i(m^A+m^B)(\delta_{ij}+\ell\epsilon_{ij}\hat{P}_1-\ell\delta_{ij}\delta_{1j}\hat{P}_2).
\end{align}
Since our ansatz for the hamiltonian is
\begin{equation}
    \hat{H}_{dS}=\hat{H}_{KIN}^{AB}+\hat{V}^{AB}_{dS}
\end{equation}
to find the potential, we impose that the commutation relations \eqref{algebracomposite1}-\eqref{algebracomposite2} are satisfied for $\hat{P}_i,\hat{R},\hat{K}_i$ and $\hat{H}_{dS}$. This results in the following conditions:
\begin{align}\label{eq: 5.110}
    &[\hat{P}_i,V^{AB}_{dS}]=0, \\ \label{eq: 5.111}
    &[\hat{R},V^{AB}_{dS}]=0, \\ \label{eq: 5.112}
    &[\hat{K}_i,V^{AB}_{dS}]=0. 
\end{align}
Enforcing conditions \eqref{eq: 5.110}-\eqref{eq: 5.112}, one finds that the proper dS composition law is admissible if the harmonic potential is deformed to
\begin{align} \nonumber
    &\hat{V}^{AB}_{dS}
    =\frac{1}{2}g(\hat{\vec{x}}^A(0)-\hat{\vec{x}}^B(0))^2+\frac{\ell g}{2}((\hat{p}_1^A-\hat{p}_1^B)(\hat{x}_1^A(0)\hat{x}_2^B(0)-\hat{x}_2^A(0)\hat{x}_1^B(0))\\ &+(\hat{x}_1^A(0)\hat{x}_2^B(0)-\hat{x}_2^A(0)\hat{x}_1^B(0))(\hat{p}_1^A-\hat{p}_1^B)).
\end{align}
We now explore the method inspired by the coproduct law. According to the galilean coalgebric sector \eqref{HP}-\eqref{KiPj} we should complete equations \eqref{frattfabCOP1}-\eqref{frattfabCOP3} with
\begin{align}
 &\hat{K}_1=\hat{k}_1^A+\hat{k}_1^B+\ell\hat{p}_1^A\hat{k}_2^B, \\
    &\hat{K}_2=\hat{k}_2^A+\hat{k}_2^B-\ell\hat{p}_1^A\hat{k}_1^B, \\
    &\hat{H}=\hat{h}^A+\hat{h}^B.
\end{align}
However, these composition laws do not satisfy the compatibility conditions \eqref{algebracomposite1}-\eqref{algebracomposite2}. In particular, one has
\begin{align}
    &[\hat{R},\hat{K}_2]=-i\hat{K}_1+i\ell m^A\hat{r}^B,\\ 
    &[\hat{K}_1,\hat{P}_1]=i(m^A+m^B-\ell(m^A+m^B)\hat{p}_2^A-\ell m^B\hat{p}_2^B).
\end{align}
As discussed at the end of \hyperref[sec: TCK Galilean limit]{Section~\ref{sec: TCK Galilean limit}}, the abstract Hopf symmetry structure $\mathcal{G}_\kappa$ does not, by itself, possess a direct physical interpretation. Furthermore, the above construction of the composition law is merely heuristic and therefore does not uniquely determine its form. This freedom allows us to deform the composition law so as to recover the physically relevant canonical realization $\mathcal{CG}_\kappa$ of equations \eqref{algebracomposite1}-\eqref{algebracomposite2}.
The compatibility is indeed restored if we change the coproduct of $\hat{K_1}$ adding a term proportional to the mass of the first particle in the following way:
\begin{align}
\hat{K}_1=\hat{k}_1^A+\hat{k}_1^B+\ell\hat{p}_1^A\hat{k}_2^B-\ell m^A\hat{r}^B. 
\end{align}
Following the same logic of the previous paragraph we make the ansatz 
\begin{equation}
\hat{H}_{\kappa}=\hat{H}_{KIN}^{AB}+\hat{V}^{AB}_{\kappa}
\end{equation}
and impose the compatibility conditions \eqref{algebracomposite1}-\eqref{algebracomposite2} for $\hat{P}_i,\hat{R},\hat{K}_i$ and $\hat{H}_{\kappa}$, leading to
\begin{align} \label{eq: 5.132}
    &[\hat{P}_i,V^{AB}_{\kappa}]=0, \\ \label{eq: 5.133}
    &[\hat{R},V^{AB}_{\kappa}]=0, \\ \label{eq: 5.134}
    &[\hat{K}_i,V^{AB}_{\kappa}]=0.
\end{align}
Enforcing conditions \eqref{eq: 5.132}-\eqref{eq: 5.134}, one finds that the coproduct composition law is indeed admissible if the potential is deformed in the following way:
\begin{align} \nonumber
    &\hat{V}^{AB}_{\kappa}=\frac{1}{2}g(\hat{\vec{x}}^A(0)-\hat{\vec{x}}^B(0))^2+\ell g(\frac{\hat{x}_2^A(0)}{2}-\frac{\hat{x}_2^B(0)}{2}+\\&+\hat{p}_2^B(\hat{x}_1^B(0))^2-\hat{p}_2^B\hat{x}_1^A(0)\hat{x}_1^B(0)+\hat{p}_1^B\circ \hat{x}_1^B(0)\hat{x}_2^A(0)-\hat{p}_1^B\circ \hat{x}_1^B(0)\circ\hat{x}_2^B(0)).
\end{align}

\section*{Conclusions and Outlook}

In the last decades, deformed relativistic symmetries have attracted a lot of interest within the quantum-gravity community, due to their intuitive appeal and potential phenomenological relevance. However, it has proven very difficult to translate their natural mathematical description in terms of quantum groups acting on noncommutative spacetime manifolds into meaningful phenomenological predictions. The first Noether theorem, which provides a link between the formal structure of ordinary Poincaré symmetries and some observable quantities, is not applicable to noncommutative spacetime symmetries, making it impossible to work out their physical consequences based on na\"ive extrapolation from commutative-spacetime intuition. While there have been a few attempts at obtaining a noncommutative generalization of the Noether theorem in the context of free Lagrangian field theory~\cite{Generalizing the Noether theorem for Hopf-algebra spacetime symmetries, nopureboost, fieldtheoryfreidel, Twisted Hopf symmetries of canonical noncommutative spacetimes and the no-pure-boost principle}, their phenomenological relevance is dubious, as the models explored in these works are dynamically trivial and undermined by serious interpretive issues. 

In this paper, inspired by the pioneering analysis reported in Ref.~\cite{fuzzy1}, we have shown that noncommutative spacetime symmetries can be consistently represented and transparently related to the structure of the algebra of the observables in a CQM-based, first-quantized setting. In particular, we have found that the usual physical consequences of the Noether theorem do still hold in the noncommutative case, but the relevant symmetry algebra is not the abstract Poincaré Hopf algebra $\mathcal{P}$, \emph{i.e.}, the natural mathematical generalization of the usual Poincaré Lie algebra $\mathcal{P}_0$, but rather a canonical symmetry algebra $\mathcal{CP}$ embedded in the extended canonical algebra of CQM, whose generators are related to the abstract ones via equations~\eqref{tcommf}-\eqref{mcommf}. When expressed in terms of the canonical generators, which are standard first-quantized observables with a straightforward physical meaning, infinitesimal symmetry transformations have the same form \eqref{dvexplicit} as in standard quantum mechanics, and all the interpretive difficulties affecting the field-theoretic approach of Refs.~\cite{Generalizing the Noether theorem for Hopf-algebra spacetime symmetries, nopureboost, fieldtheoryfreidel, Twisted Hopf symmetries of canonical noncommutative spacetimes and the no-pure-boost principle}, such as the puzzling ``no-pure boost'' phenomenon first noticed in Ref.~\cite{nopureboost}, disappear. While our results are conditional on the availability of a representation of the type~\eqref{Epsilon}-\eqref{Omega} for the Poincaré noncommutative transformation parameters, they apply to most, if not all, noncommutative spacetimes of physical interest, providing a rather broad generalization of the Noether theorem to noncommutative spacetime symmetries. 

Focusing on the time‑commutative $\kappa$-plane as a concrete case study, we have then shown step by step how to introduce spacetime noncommutativity into the extended canonical algebra of CQM, how to find a representation of the canonical generators in terms of the extended canonical variables, and how to derive the commutation structure of the relevant canonical symmetry algebra. In order to do this, we first had to fill a gap in the existing literature and find the complete TC$\kappa$ Poincaré Hopf algebra $\mathcal{P}_\kappa$, including boosts and time translations, at first order in the noncommutativity scale $\ell$. Our leading-order approach, from the characterization of the abstract Poincaré Hopf algebra and the associated noncommutative transformation parameters to the determination of the corresponding first-quantized canonical symmetry algebra, can be generalized to most, if not all, Lie-algebra spacetimes, resulting in a wealth of first-quantized noncommutative spacetime models whose free-particle phenomenology can be rigorously and quantitatively characterized along the lines of Ref.~\cite{fuzzy2}. A systematic investigation of this promising theoretical landscape, at both the abstract and the canonical levels, is currently in progress, and will be the subject of a forthcoming study.

In the second part of this manuscript, following a suggestion first advanced in Ref.~\cite{soccerball} and heuristically developed in Ref.~\cite{frattfab}, we have focused our attention on the Galilean-relativistic limit of our TC$\kappa$ model. Thanks to the commutativity of the time coordinate, we have been able to characterize the single-particle algebra of the observables $\mathcal{O}_\kappa$ as a deformed Heisenberg algebra, identifying the noncommutative analogues $\hat{x}_j(0)$ of the initial-time Heisenberg position observables of standard quantum mechanics and determining their action as differential operators on the momentum representation space $L^2(\mathbb{R}^2,d^2p)$. In this way, we have obtained an ordinary, non-covariant (but fully relativistic), first-quantized description of a free TC$\kappa$ particle, in which time evolution is encoded in a deformation of the usual Galilean-relativistic Hamiltonian $\hat{p}_k\hat{p}_k/2m$. Starting from such a description, it is possible to build models of interacting TC$\kappa$ particles in the usual way, \emph{i.e.}, by adding an interaction potential to the two-particle TC$\kappa$ Hamiltonian, and explore the crucial interplay between deformed relativistic symmetries and dynamics, which has eluded all previous field-theoretic efforts.

As a first exercise in this line of research motivated by the pioneering analysis reported in Ref.~\cite{frattfab}, in the last section we have analyzed a two‑particle system interacting via a deformed harmonic potential. We have found that, unlike the standard case, the expressions of the two-particle canonical generators in terms of the single-particle observables are not uniquely determined independently of the dynamics. Rather, it appears that there are several admissible composition laws for the symmetry generators and that they are closely associated to the form of the interaction potential, substantiating the heuristic results obtained in Ref.~\cite{frattfab} for purely spatial symmetries. With this admittedly preliminary and incomplete analysis, we have only scratched the surface of the issue of two-particle generators in the presence of noncommutative spacetime symmetries. We feel that our first-quantized approach could provide much deeper insights into this significant theoretical challenge. For example, a systematic classification of the admissible composition laws, for generic time-commutative spacetimes, could hopefully lead to the identification of some physically distinguished ones, which could inform future searches of phenomenological signatures of deformed relativistic symmetries. Generalizing our framework to more realistic 3+1D scenarios, in particular extending it to particle with non-zero spin, could also enable a thorough investigation of angular-momentum physics in rotationally invariant systems, like the hydrogen atom, opening intriguing phenomenological windows into spacetime noncommutativity in the Galilean-relativistic limit.


\begin{thebibliography}{99}

    \bibitem{qstph}
    G. Amelino-Camelia, \emph{Quantum-spacetime phenomenology}, Living Rev. Relativ. \textbf{16} (2013), 5. 

    \bibitem{Hopf algebras for physics at the Planck scale}
    S. Majid,
    \emph{Hopf algebras for physics at the Planck scale},
     Class. Quantum Grav. \textbf{5} (1988), 1587.   

    \bibitem{Drinfeld} 
    V. G. Drinfeld, \emph{Quantum groups}, J. Math. Sci. \textbf{41} (1988), 898. 

    \bibitem{foundofquangro}
    S. Majid, \emph{Foundations of quantum group theory}, Cambridge University Press (2000).

    \bibitem{Quantum Groups and Noncommutative Geometry}
    S. Majid, \emph{Quantum groups and noncommutative geometry},
    J. Math. Phys \textbf{41} (2000), 3892. 

    \bibitem{waves}
    G. Amelino-Camelia and S. Majid, \emph{Waves on noncommutative
    space–time and gamma-ray bursts}, Int. J. Mod. Phys. A \textbf{15} (2000),
    4301. 

    \bibitem{Quantum Field Theory on Noncommutative Spaces}
    R. Szabo, \emph{Quantum field theory on noncommutative spaces}, Phys. Rept. \textbf{378} (2003), 207. 

    \bibitem{Generalizing the Noether theorem for Hopf-algebra spacetime symmetries} 
    A. Agostini, G. Amelino-Camelia, M. Arzano, A. Marcianó and R. Altair Tacchi, \emph{Generalizing the Noether theorem for Hopf-algebra spacetime symmetries}, Mod. Phys. Lett \textbf{22} (2007), 1779. 

    \bibitem{nopureboost}
    G. Amelino-Camelia, G. Gubitosi, A. Marcianò, P. Martinetti and F. Mercati, \emph{A no-pure-boost uncertainty principle from spacetime noncommutativity}, 	Phys.Lett. B \textbf{671}        (2009), 298. 

    \bibitem{fieldtheoryfreidel}
   L. Freidel, J. Kowalski-Glikman and S. Nowak,
   \emph{Field theory on $\kappa$-minkowski space revisited: Noether charges and breaking of lorentz symmetry}, Int. J. Mod. Phys. A \textbf{23} (2008), 2687. 

   \bibitem{Twisted Hopf symmetries of canonical noncommutative spacetimes and the no-pure-boost principle}
   G. Amelino-Camelia, F. Briscese, G. Gubitosi, A. Marcianò, P. Martinetti and F. Mercati, \emph{Twisted Hopf symmetries of canonical noncommutative spacetimes and the no-pure-boost principle}, Phys. Rev. D \textbf{78} (2008), 025005. 

   \bibitem{moia}
   A. Moia, \emph{Noncommutative spacetime symmetries from covariant quantum mechanics}  Adv. High Energy Phys. \textbf{2017} (2017), 4042314. 

   \bibitem{soccerball}
    G. Amelino-Camelia, \emph{Planck-scale soccer-ball problem: A case of mistaken
    identity}, Entropy \textbf{19} (2017), 400. 

    \bibitem{frattfab}
    G. Amelino-Camelia, G. Fabiano and D. Frattulillo, \emph{Total momentum and other noether charges for particles interacting in a quantum spacetime}, Symmetry \textbf{17} (2025), 227. 

    \bibitem{bicross}
    S. Majid and H. Ruegg, \emph{Bicrossproduct structure of the k-Poincare group and non-commutative geometry}, Phys. Lett. B. \textbf{334} (1994), 348. 

    \bibitem{Classical and Quantum Mechanics of Free  Relativistic Systems}
    J. Lukierski, H. Ruegg and W.J. Zakrzewski, \emph{Classical and Quantum Mechanics of Free  Relativistic Systems}, Ann. Phys. \textbf{243} (1995), 90. 

     \bibitem{New quantum Poincaré algebra and κ-deformed field theory} 
     J. Lukierski, A. Nowicki and H. Ruegg, \emph{New quantum Poincaré algebra and $\kappa$-deformed field theory}, Phys. Lett. B \textbf{293} (1992), 344. 

     \bibitem{ballentine}
     L.E. Ballentine, \emph{Quantum mechanics: a modern development}, World Scientific (1998).

     \bibitem{Trajectories for the wave function of the universe from a simple detector model}
     J. Halliwell, \emph{Trajectories for the wave function of the universe from a simple detector model}, Phys. Rev. D \textbf{64} (2001), 044008. 


     \bibitem{Relational time in generally covariant quantum systems: Four models} 
     R. Gambini and R. A. Porto, \emph{Relational time in generally covariant quantum systems: Four models}, Phys. Rev. D \textbf{63} (2001), 105014. 

    \bibitem{Spacetime states and covariant quantum theory}
    M. Reisenberger and C. Rovelli, \emph{Spacetime states and covariant quantum theory}, Phys. Rev. D \textbf{65} (2002), 125016. 

    \bibitem{quantization of Gauge Systems}
    M.Henneaux and C. Teitelboim, \emph{Quantization of Gauge Systems}, Princeton University Press (1992). 

    \bibitem{Refined algebraic quantization: Systems with a single constraint}
    D. Marolf, \emph{Refined algebraic quantization: Systems with a single constraint}, Banach Center Publications \textbf{39} (1997), 331.

    \bibitem{fuzzy1}
    G. Amelino-Camelia, V. Astuti and G. Rosati, \emph{Relative locality in a quantum spacetime and the pregeometry of $\kappa$-minkowski}, Eur. Phys. J. C \textbf{73} (2013), 2521. 

    \bibitem{fuzzy2}
    G. Amelino-Camelia, Valerio Astuti and G. Rosati, \emph{Predictive description of planck-scale-induced spacetime fuzziness}, Phys. Rev. D \textbf{87} (2013), 084023.

    \bibitem{woronowicz}
    S.L. Woronowicz, \emph{Differential calculus on compact matrix pseudogroups (quantum groups)}, Commun.Math. Phys. \textbf{122} (1989), 125 .

    \bibitem{jordancurriesudarshan}
    D.G. Currie, T.F. Jordan, and E.C.G. Sudarshan, \emph{Relativistic invariance and hamiltonian theories of interacting particles}, Rev. Mod. Phys. \textbf{35} (1963), 350. 

    \bibitem{qgprimer}
    S. Majid, \emph{A Quantum Groups Primer}, Cambridge University Press, 2002.

    \bibitem{tminkowski}
    F. Mercati, \emph{T-minkowski noncommutative spacetimes I: Poincaré groups, differential calculi, and braiding}, Prog. Theor. Exp. Phys. \textbf{2024} (2024), 073B06.

    \bibitem{strocchiqft}
    F. Strocchi, \emph{An Introduction to Non-Perturbative Foundations of Quantum Field Theory}, International Series of Monographs on Physics, Oxford University Press (2013).

    \bibitem{strocchiqm}
    F. Strocchi, \emph{An Introduction to the Mathematical Foundation of Quantum Mechanics}, World Scientific (2005).

    \bibitem{desitter}
    G.~Amelino-Camelia, G.~Gubitosi and G.~Palmisano,
    \emph{Pathways to relativistic curved momentum spaces: de Sitter case study},
    Int. J. Mod. Phys. D \textbf{25} (2016), 1650027.

\end{thebibliography}
\end{document}